\documentclass[letterpaper, 10 pt, journal, twoside]{IEEEtran}	% Use this command for final version

\IEEEoverridecommandlockouts                              % This command is only needed if 
\usepackage{graphicx}
\usepackage[utf8]{inputenc}
\usepackage{psfrag}
\usepackage{amsmath}
\usepackage{amssymb}
\usepackage{xcolor}	
\usepackage{cite}
\usepackage{subfigure}

\usepackage{multirow}

\newtheorem{assumption}{Assumption}

\newcommand{\paper}{letter}

\title{Analysis and control of genetic toggle switches subject to periodic multi-input stimulation}

\author{Davide Fiore$^{1}$, Agostino Guarino$^{1}$, Mario di Bernardo$^{1,2}$% <-this % stops a space
\thanks{$^{1}$Davide Fiore, Agostino Guarino and Mario di Bernardo are with the Department of Electrical Engineering and Information Technology, University of Naples Federico II, Via Claudio 21, 80125 Naples, Italy.
        {\tt\small dvd.fiore@gmail.com, agostinoguarino@gmail.com}}%
\thanks{$^{2}$Mario di Bernardo is also with the Department of Engineering Mathematics, University of Bristol, University Walk, BS8 1TR Bristol, U.K.
        {\tt\small mario.dibernardo@unina.it}}%
}

\begin{document}

\maketitle
\pagestyle{empty}
\thispagestyle{empty}

%%%%%%%%%%%%%%%%%%%%%%%%%%%%%%%%%%%%%%%%%%%%%%%%%%%%%%%%%%%%%%%%%%%%%%%%%%%%%%%%
\begin{abstract}
In this \paper, we analyze a genetic toggle switch recently studied in the literature where the expression of two repressor proteins can be tuned by controlling two different inputs, namely the concentration of two inducer molecules in the growth medium of the cells. Specifically, we investigate the dynamics of this system when subject to pulse-width modulated (PWM) input. We provide an analytical model that captures qualitatively the experimental observations reported in the literature and approximates its asymptotic behavior. We also discuss the effect that the system parameters have on the prediction accuracy of the model. Moreover, we propose a possible external control strategy to regulate the mean value of the fluorescence of the reporter proteins when the cells are subject to such periodic forcing.
\end{abstract}

% Keywords appear just beneath the abstract. Use only for final version.
\begin{IEEEkeywords}
Systems biology, Genetic regulatory systems, Modeling
\end{IEEEkeywords}

%%%%%%%%%%%%%%%%%%%%%%%%%%%%%%%%%%%%%%%%%%%%%%%%%%%%%%%%%%%%%%%%%%%%%%%%%%%%%%%%
%
\section{Introduction}
\IEEEPARstart{T}{he} genetic toggle switch is a fundamental component in synthetic biology as it plays a major role in cell differentiation and decision making \cite{alon2006introduction,chen2010modeling}. Its importance comes from its ability to endow host cells with memory of some previous stimulus reporting this information as high expression rate of a specific repressor protein \cite{gardner2000construction,tian2006stochastic,wu2013engineering}. 

The genetic toggle switch as first designed in \cite{gardner2000construction} consists of two repressor proteins, both repressing each other's promoter, so that only one protein is fully expressed at any time. From a modelling viewpoint, the genetic toggle switch is a bistable dynamical system, possessing two stable equilibria, each associated to a fully expressed protein, and a saddle equilibrium point, whose stable manifold is the boundary separating the basins of attraction of the other two.

Different approaches have been presented to control the response of genetic toggles switches. Examples include methods based on piecewise affine approximations \cite{chaves2011exact}, pulse shaping of the external inputs based on monotone systems theory \cite{sootla2016shaping}, and the analysis of the stationary probability distributions of the outputs in different working conditions \cite{petrides2017understanding}.

\begin{figure}[!t]
\begin{center}
\includegraphics[width=0.7\columnwidth]{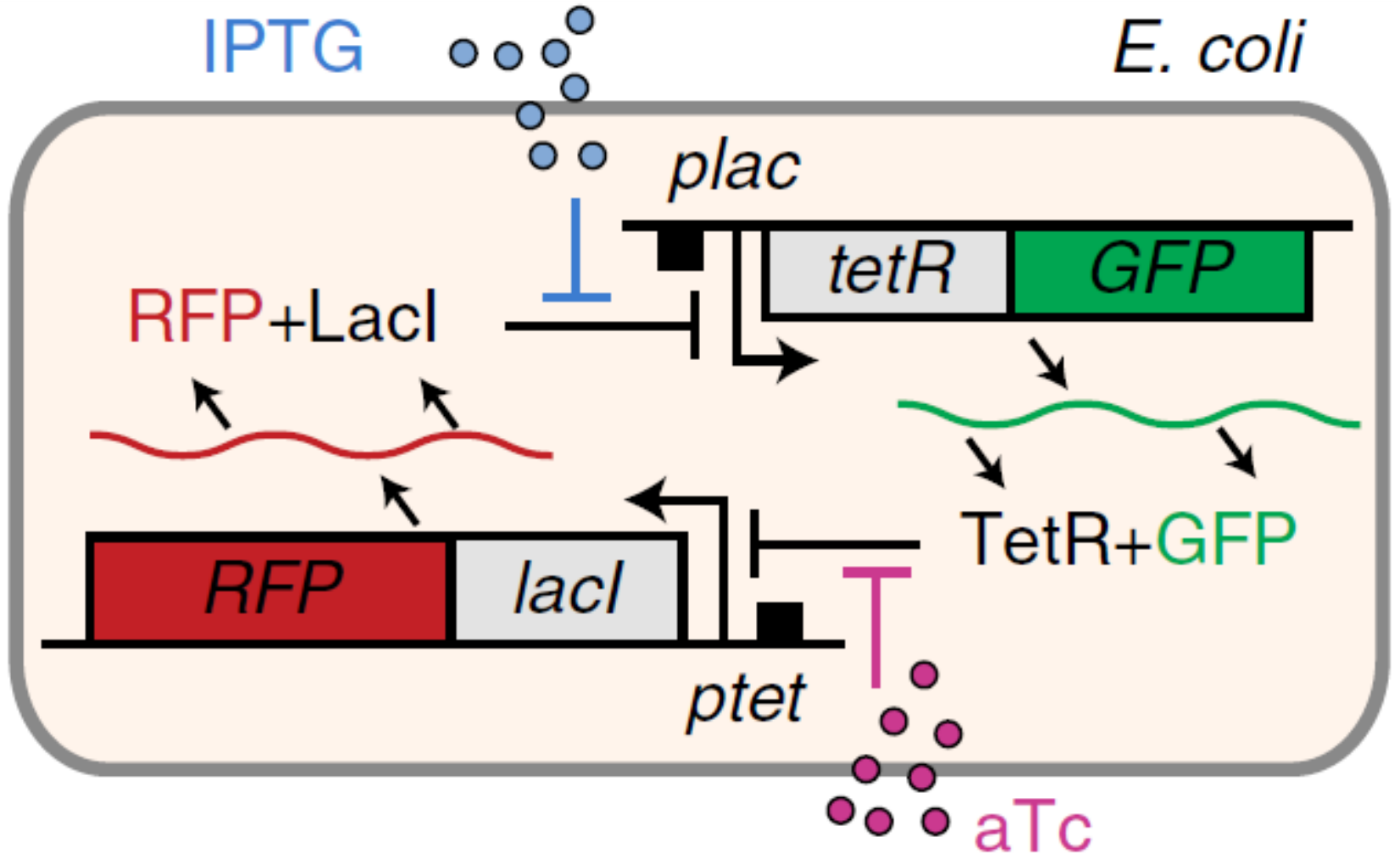}
\caption{Genetic toggle switch embedded in \emph{E.coli} considered in \cite{lugagne2017balancing} (Figure reused under Creative Commons license).}
\label{fig:ecoli_ts}
\end{center}
\end{figure}

Recently, in \cite{lugagne2017balancing} the problem has been studied of dynamically ``balancing'' a genetic toggle switch (based on the LacI/TetR promoters in {E.coli}, schematically shown in Figure \ref{fig:ecoli_ts}) in an undecided state somewhere in between its two stable equilibrium points. The expression level of the two repressing proteins can be controlled by regulating the concentration of two inducer molecules, aTc and IPTG.
The former, aTc, binds to TetR, increasing the rate of production of LacI, and therefore causing the cell to commit to the stable equilibrium point corresponding to high expression of LacI (high LacI/low TetR). The latter, IPTG, binds instead to LacI, causing the commitment of the cell to the other stable equilibrium point (high TetR/low LacI).  From a dynamical systems viewpoint,  varying the two input signals causes the occurrence of two saddle-node bifurcations changing the phase portrait of the system from bistability to monostability (Figure \ref{fig:nullclines}). 

In their work Lugagne et al. \cite{lugagne2017balancing} focus on both the problem of controlling a single cell and that of taming the behavior of the whole population. Their approach is based on considering the toggle switch as a multi-input control system and is aimed at using both inputs to keep the switch evolving in a neighborhood of its saddle point; a problem they propose as a test-bed scenario in synthetic biology similar to that of stabilizing an inverted pendulum in classical control.

When implementing single cell control, the fluorescence level of the reporter proteins in a single cell are measured and compared to their reference values. Two different classes of controllers were used in \cite{lugagne2017balancing}, PI and bang-bang, both designed independently for each control input (aTc and IPTG). Using PI controllers on both input channels, it is possible to make the single cell evolve (oscillate) near the saddle point. Although the controlled cell follows (on average) the desired reference, the rest of the population is observed to drift away, converging instead to some other equilibrium point. 

Surprisingly, it is reported in \cite{lugagne2017balancing} that this undesired effect is absent when the \emph{single} cell is controlled by two independent bang-bang inputs with the rest of the population exhibiting an evolution similar to the target cell in this case. 
To further explore this effect, the authors then consider an open-loop \emph{periodic stimulation} (two mutually exclusive pulse waves with prescribed width) to control the whole population. Again the whole population is shown to converge to some periodic orbit surrounding the saddle point with a remarkable level of coherence in terms of both mean and standard deviation despite cell-to-cell variability and other phenotypic differences between cells.

Using an in-silico model this effect is explained in \cite{lugagne2017balancing} as due to the phase portrait of the forced system periodically changing from one presenting a unique high-LacI equilibrium point to another with a unique high-TetR equilibrium point.
Heuristically, this results in an \emph{average} phase portrait having a unique attractor in between the former two given that, as conjectured in \cite{lugagne2017balancing}, the cell dynamics and the periodic excitation act on different time-scales. Also, changing the characteristics of the periodic PWM forcing (such as period, width and amplitude of the pulses) shifts the position of the average attractor causing cells to evolve towards a different target solution.

Despite providing some qualitative explanation of the experimental observations, several open questions remain. For instance, what causes the massive reduction in standard deviation between different cells in the population and what the period/duty cycle should be of the control inputs to achieve the desired behavior. Also, the challenge remains of designing better multi-input {\em feedback} strategies to control populations of host cells endowed with synthetic toggle switches. 

In this \paper, we address some of these open problems by providing an analytical investigation of the phenomena reported in \cite{lugagne2017balancing}. We start by deriving a \emph{quasi-steady state model} of the toggle-switch system proposed therein. Using formal averaging techniques for nonlinear systems \cite{khalil2002nonlinear}, we derive an autonomous \emph{average vector field}, whose solutions, under some conditions, approximate those of the original time-varying system. 
To simplify the analysis, we assume that the diffusion of the inducer molecules across the cell membrane is \emph{instantaneous}. 

We prove that if the average vector field has a unique attracting equilibrium point $\bar{x}_\mathrm{av}$, whose position in  state space depends on the duty cycle $D$ and on the amplitude of the forcing pulse waves $u_{\mathrm{aTc}}(t)$ and $u_{\mathrm{IPTG}}(t)$, then every solution of the original time-varying system asymptotically converges to a periodic orbit in some neighborhood of $\bar{x}_\mathrm{av}$. We compare our model predictions with the experimental observations made in \cite{lugagne2017balancing} and with the mean-value trajectories of the original model proposed therein.  We use the model and its analysis to provide some indications on how the parameters of the toggle switch may be tuned to enhance its response to the class of periodic inputs of interest, and exploit the results to synthesize an external control strategy to regulate the mean-value of the measured fluorescence of the reporter proteins in the cell at some desired value.
We wish to emphasize that the analysis provided in this {\paper} can be instrumental for the design of further control strategies for this particularly relevant class of synthetic devices and to investigate the effects at the population level of different types of periodic stimuli to the cells.
%
%
%%%%%%%%%%%%%%%%%%%%%%%%%%%%%%%%%%%%%%%%%%%%%%%%%%%%%%%%%%%%%%%%%%%%%%%%
%
%
\section{Mathematical model of the toggle switch}
\label{sec:model_and_input}
\subsection{Transcription-translation model}
The deterministic model of the toggle switch that we start from can be given as follows \cite{lugagne2017balancing}

%\vspace{-0.6cm}
\footnotesize
\begin{align}
\label{eq:transcr_laci}
& \begin{aligned}
\frac{d\, mRNA_{\mathrm{LacI}}}{dt}=\; & \kappa_\mathrm{L}^\mathrm{m0} + \frac{\kappa_\mathrm{L}^\mathrm{m}}{1+ \left( \frac{TetR}{\theta_{\mathrm{TetR}} } \cdot \frac{1}{1 + \left( aTc/\theta_{\mathrm{aTC}} \right)^{\eta_{\mathrm{aTc}}} } \right)^{\eta_{\mathrm{TetR}}} } \\
 & - g_\mathrm{L}^\mathrm{m} \cdot mRNA_{\mathrm{LacI}}
\end{aligned}
\\
\label{eq:trascr_tetr}
& \begin{aligned}
\frac{d\, mRNA_{\mathrm{TetR}}}{dt}=\; & \kappa_\mathrm{T}^\mathrm{m0} + \frac{\kappa_\mathrm{T}^\mathrm{m}}{1+ \left( \frac{LacI}{\theta_{\mathrm{LacI}} } \cdot \frac{1}{1 + \left( IPTG/\theta_{\mathrm{IPTG}} \right)^{\eta_{\mathrm{IPTG}}} } \right)^{\eta_{\mathrm{LacI}}} } \\
 & - g_\mathrm{T}^\mathrm{m} \cdot mRNA_{\mathrm{TetR}} 
\end{aligned}
\\
\label{eq:transl_laci}
& \frac{d\, LacI}{dt}= \kappa_\mathrm{L}^\mathrm{p} \cdot mRNA_{\mathrm{LacI}} - g_\mathrm{L}^\mathrm{p} \cdot LacI\\
\label{eq:transl_tetr}
& \frac{d\, TetR}{dt}= \kappa_\mathrm{T}^\mathrm{p} \cdot mRNA_{\mathrm{TetR}} - g_\mathrm{T}^\mathrm{p} \cdot TetR
\end{align}
\normalsize
In the above equations the variables denote concentrations of molecules inside the cell, and the parameters $\kappa_\mathrm{L/T}^\mathrm{m0}$, $\kappa_\mathrm{L/T}^\mathrm{m}$, $\kappa_\mathrm{L/T}^\mathrm{p}$, $g_\mathrm{L/T}^\mathrm{m}$, $g_\mathrm{L/T}^\mathrm{p}$ are leakage transcription, transcription, translation, mRNA degradation, and protein degradation rates, respectively. All parameter values are provided in Supplementary Table 1 and are also the same used in \cite{lugagne2017balancing}.

The inducer molecules diffuse in and out of the cell across the membrane with non-symmetrical exchange dynamics modeled by

%\vspace{-0.3cm}
\footnotesize
\begin{align}
\label{eq:diffusion_atc}
\frac{d\, aTc}{dt}= &
\begin{cases}
k^{\mathrm{in}}_{\mathrm{aTc}} (u_{\mathrm{aTc}} - aTc), & \mbox{ if }\ u_{\mathrm{aTc}} > aTc\\
k^{\mathrm{out}}_{\mathrm{aTc}} (u_{\mathrm{aTc}} - aTc), & \mbox{ if }\ u_{\mathrm{aTc}} \leq aTc
\end{cases},\\
\label{eq:diffusion_iptg}
\frac{d\, IPTG}{dt}= &
\begin{cases}
k^{\mathrm{in}}_{\mathrm{IPTG}} (u_{\mathrm{IPTG}} - IPTG), & \mbox{ if }\ u_{\mathrm{IPTG}} > IPTG\\
k^{\mathrm{out}}_{\mathrm{IPTG}} (u_{\mathrm{IPTG}} - IPTG), & \mbox{ if }\ u_{\mathrm{IPTG}} \leq IPTG
\end{cases},
\end{align}
\normalsize 
where $aTc$ and $IPTG$ denote the concentrations of the inducer molecules inside the cell, while $u_{\mathrm{aTc}}$ and $u_{\mathrm{IPTG}}$ those in the growth medium.
\begin{figure}[!t]
\begin{center}
\includegraphics[width=0.75\columnwidth]{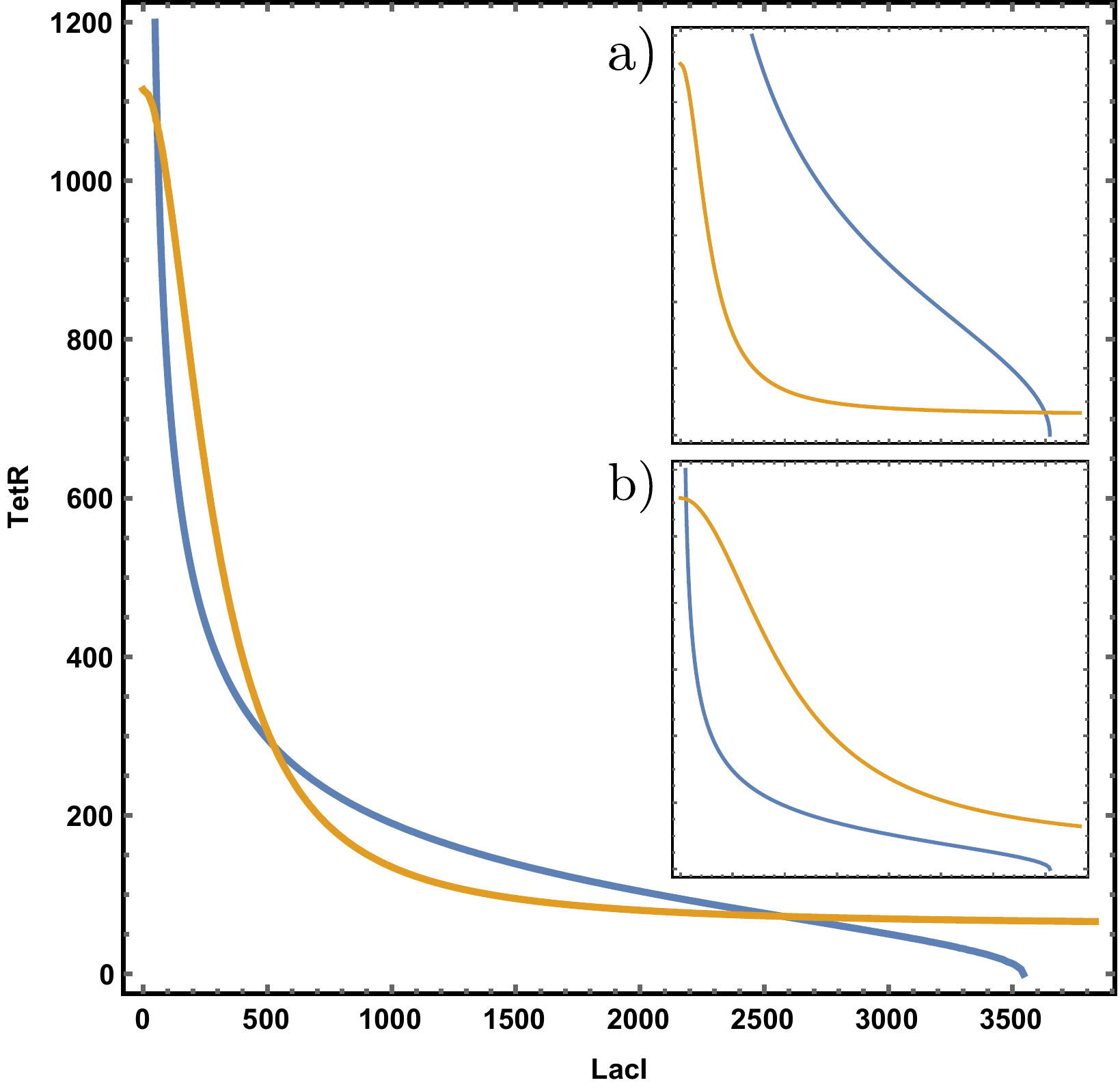}
\caption{Nullclines of the toggle switch system \eqref{eq:sys_nondim}. Main picture: bistability: two stable and one saddle equilibrium points. Reference values $aTc=20\,\mathrm{ng/ml}$, $IPTG=0.25\,\mathrm{mM}$. Insets: a) monostability: unique high LacI/low TetR equilibrium point. $aTc=50\,\mathrm{ng/ml}$, $IPTG=0.25\,\mathrm{mM}$; b) monostability: unique high TetR/low LacI equilibrium point. $aTc=20\,\mathrm{ng/ml}$, $IPTG=0.50\,\mathrm{mM}$ }
\label{fig:nullclines}
\end{center}
\end{figure}
\subsection{Quasi-steady state model}
Assuming that the concentrations of the mRNA molecules reach steady state more rapidly than their corresponding proteins, that LacI and TetR proteins degrade at the same rate, that is $g_\mathrm{L}^\mathrm{p}=g_\mathrm{T}^\mathrm{p}=g^\mathrm{p}$, and using the following dimensionless variables (similarly as done in \cite{kuznetsov2004synchrony,nikolaev2016quorum})
\begin{equation}
\label{eq:adim_variables}
t'=g^\mathrm{p}\, t, \ \ x_1=\frac{LacI}{\theta_{\mathrm{LacI}} }, \ \ x_2=\frac{TetR}{\theta_{\mathrm{TetR}}},
\end{equation}
we obtain the following nondimensional quasi-steady state model of the genetic toggle switch
\begin{equation}
\label{eq:sys_nondim}
\begin{split}
\frac{dx_1}{dt'} &= k_1^0 + \frac{k_1}{1+ x_2^2 \cdot w_1(t'/g^\mathrm{p}) }  - x_1\\
\frac{dx_2}{dt'} &= k_2^0 + \frac{k_2}{1+ x_1^2 \cdot w_2(t'/g^\mathrm{p}) }  - x_2
\end{split}
\end{equation}
where
\begin{equation}
\label{eq:parameter_adim_1}
k_1^0=\frac{\kappa_\mathrm{L}^\mathrm{m0}\,\kappa_\mathrm{L}^\mathrm{p} }{g_\mathrm{L}^\mathrm{m}\,\theta_{\mathrm{LacI}}\, g^\mathrm{p}  }, \quad k_1=\frac{ \kappa_\mathrm{L}^\mathrm{m}\,\kappa_\mathrm{L}^\mathrm{p}}{g_\mathrm{L}^\mathrm{m}\,\theta_{\mathrm{LacI}}\, g^\mathrm{p}  },
\end{equation}
and
\begin{equation}
\label{eq:parameter_adim_2}
k_2^0=\frac{\kappa_\mathrm{T}^\mathrm{m0}\,\kappa_\mathrm{T}^\mathrm{p} }{g_\mathrm{T}^\mathrm{m}\,\theta_{\mathrm{TetR}}\, g^\mathrm{p}  }, \quad k_2=\frac{\kappa_\mathrm{T}^\mathrm{m}\,\kappa_\mathrm{T}^\mathrm{p} }{g_\mathrm{T}^\mathrm{m}\,\theta_{\mathrm{TetR}}\, g^\mathrm{p}  },
\end{equation}
are dimensionless parameters, and we have set $\eta_{\mathrm{LacI}}=\eta_{\mathrm{TetR}}=2$.
The steps of the previous derivation are reported in the Supplementary Material.

\begin{figure}[!t]
\begin{center}
\includegraphics[width=\columnwidth]{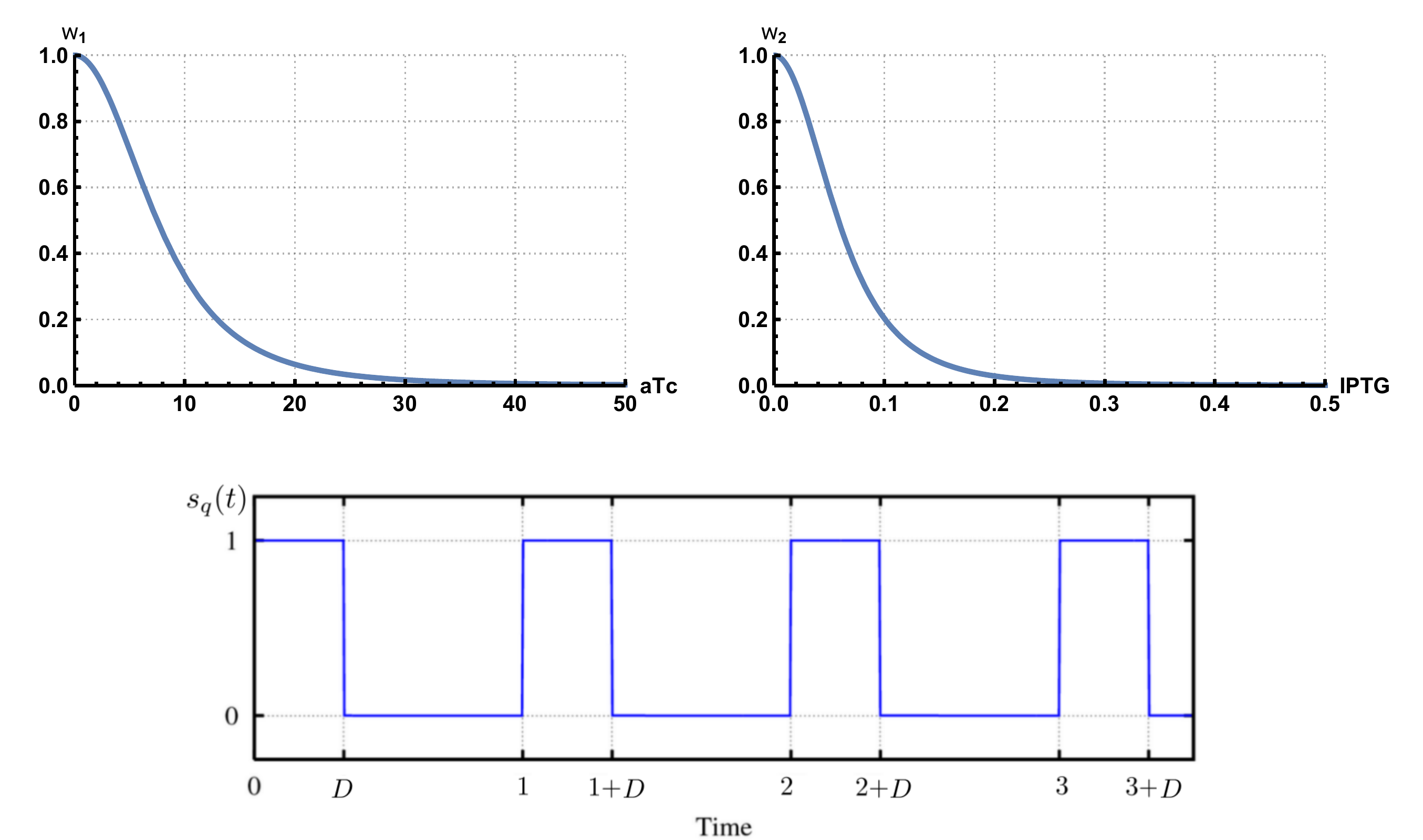}
\caption{Top: Static nonlinear functions $w_1(aTc)$ and $w_2(IPTG)$ as in \eqref{eq:w1_function} and \eqref{eq:w2_function}. Bottom: Pulse wave $s_\mathrm{q}(t)$: period $1$, duty cycle $D\in[0,1]$.}
\label{fig:w_functions_sq}
\end{center}
\end{figure}

The nonlinear functions $w_1(t)$ and $w_2(t)$ in \eqref{eq:sys_nondim}  take into account the static relationship between the repressor protein (TetR or LacI) and their regulator molecule (aTc or IPTG, respectively). They are shown in Figure  \ref{fig:w_functions_sq} and are defined as
\begin{align}
\label{eq:w1_function}
w_1(aTc(t))= &  \frac{1}{\left(1 + \left( \frac{aTc(t)}{\theta_{\mathrm{aTC}}} \right)^{\eta_{\mathrm{aTc}}} \right)^{\eta_{\mathrm{TetR}}} } \\
\label{eq:w2_function}
w_2(IPTG(t))= & \frac{1}{\left(1 + \left( \frac{IPTG(t)}{\theta_{\mathrm{IPTG}}} \right)^{\eta_{\mathrm{IPTG}}} \right)^{\eta_{\mathrm{LacI}}}}
\end{align}

\begin{figure}[!t]
\begin{center}
\includegraphics[width=\columnwidth]{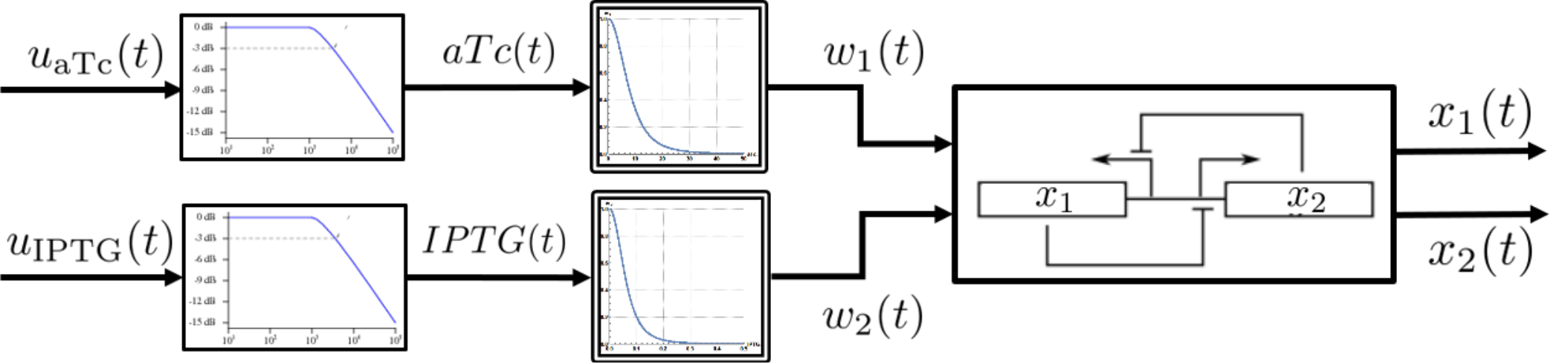}
\caption{Block diagram of system \eqref{eq:sys_nondim} with diffusion dynamics across the cell membrane \eqref{eq:diffusion_atc}-\eqref{eq:diffusion_iptg}.}
\label{fig:block_plant}
\end{center}
\end{figure}

System \eqref{eq:sys_nondim} with the static relations \eqref{eq:w1_function}-\eqref{eq:w2_function} and diffusion dynamics across the cell membrane \eqref{eq:diffusion_atc}-\eqref{eq:diffusion_iptg} can be represented in block form as in Figure \ref{fig:block_plant}. The cell membrane acts as a linear (non-symmetrical) first order low-pass filter for the signals $u_{\mathrm{aTc}}(t)$ and $u_{\mathrm{IPTG}}(t)$ with a cut-off frequency that depends on the diffusion exchange rates $k_{\mathrm{aTc}}^{\mathrm{in/out}}$ and $k_{\mathrm{IPTG}}^{\mathrm{in/out}}$. Hence, $aTc(t)$ and $IPTG(t)$ are filtered version of their respective input signals whose attenuation depends both on the cut-off frequency and on their spectral density.

In our analysis we make the following simplifying assumption. 
\begin{assumption}
The diffusion dynamics of the inducer molecules, aTc and IPTG, across the cell membrane is instantaneous, that is 
\label{ass:diffusion}
\begin{align}
aTc(t)&=u_{\mathrm{aTc}}(t),\\
IPTG(t)&=u_{\mathrm{IPTG}}(t),
\end{align}
for every $t\geq t_0$.
\end{assumption}

Later in Section \ref{sec:simulations}, we will compare our results derived from system \eqref{eq:sys_nondim} under the above Assumption \ref{ass:diffusion} with the solutions of the complete toggle switch model \eqref{eq:transcr_laci}-\eqref{eq:transl_tetr} with more realistic diffusion dynamics given by \eqref{eq:diffusion_atc}-\eqref{eq:diffusion_iptg}.
%
%
%%%%%%%%%%%%%%%%%%%%%%%%%%%%%%%%%%%%%%%%%%%%%%%%%%%%%%%%%%%%%%%%%%%%%%
%
%
\section{Averaging analysis of the toggle switch under PWM input signals}
\subsection{Forcing signals}
Following \cite{lugagne2017balancing}, the concentrations of the inducers in the growth medium are selected as two mutually exclusive pulse waves of period $T$, duty cycle $D\in[0,1]$ and amplitude $\bar{u}_{\mathrm{aTc}}$ and $\bar{u}_{\mathrm{IPTG}}$, respectively, that is
\begin{align}
\label{eq:u_atc_squarewave}
u_{\mathrm{aTc}}(t)&= \bar{u}_{\mathrm{aTc}} \cdot \left(1-s_\mathrm{q}\left( t/T \right) \right)\\
\label{eq:u_iptg_squarewave}
u_{\mathrm{IPTG}}(t)&= \bar{u}_{\mathrm{IPTG}} \cdot s_\mathrm{q}\left( t/T \right)
\end{align}
where $s_\mathrm{q}(t)$ is the pulse wave taking values 0 and 1, with period $1$ and duty cycle $D$, reported in Figure \ref{fig:w_functions_sq}. In the experiments described in \cite{lugagne2017balancing}, the amplitude $\bar{u}_{\mathrm{aTc}}$ and $\bar{u}_{\mathrm{IPTG}}$ were allowed to take values between $0$ and $100\,\mathrm{ng/ml}$, and $0$ and $1\, \mathrm{mM}$, respectively.\\
Note that $D=0$ corresponds to ``high aTc/no IPTG'' in the growth medium which in turns results in full steady-state expression of LacI (high $x_1$). Likewise,  $D=1$ corresponds to ``no aTc/high IPTG''  yielding full expression of TetR (high $x_2$). Therefore, the duty cycle can be used to control the ratio between the activation time of the two monostable systems associated to the presence or absence of the two inducer molecules whose nullclines are shown in the insets in Figure \ref{fig:nullclines}.

Under Assumption \ref{ass:diffusion} it follows that
\begin{equation}
\label{eq:atc_squarewave}
\begin{split}
w_1(t)& = w_1(aTc(t))
%& = w_1(u_{\mathrm{aTc}}(t))\\
 = w_1\left(  \bar{u}_{\mathrm{aTc}} \cdot \left(1-s_\mathrm{q}\left( t/T \right) \right) \right)\\
& = \bar{w}_1 + (1- \bar{w}_1) \cdot s_\mathrm{q}\left(t/T\right),
\end{split}
\end{equation}
where $\bar{w}_1=w_1( \bar{u}_{\mathrm{aTc}} )$, and 
\begin{equation}
\label{eq:iptg_squarewave}
\begin{split}
w_2(t)& = w_2(IPTG(t))
%& = w_2(u_{\mathrm{IPTG}}(t))\\
 = w_2\left(  \bar{u}_{\mathrm{IPTG}} \cdot s_\mathrm{q}\left( t/T \right)  \right)\\
%& = \bar{w}_2 + (1- \bar{w}_2) \cdot \left( 1- s_\mathrm{q}\left(t/T\right) \right),
& = 1 - (1-\bar{w}_2) \cdot s_\mathrm{q}\left(t/T\right),
\end{split}
\end{equation}
where $\bar{w}_2=w_2( \bar{u}_{\mathrm{IPTG}} )$. 
Therefore, $w_i(t)$ is a pulse wave taking values between $1$ and $\bar{w}_i$.
\subsection{Average vector field}
By rescaling time setting $\tau=\frac{t'}{T g^\mathrm{p}}$, system \eqref{eq:sys_nondim} can be recast as
\begin{equation}
\label{eq:sys_orig}
\begin{split}
\frac{d x_1}{d\tau} & =  \varepsilon \left[ k_1^0 + \frac{k_1}{1+ x_2^2 \cdot w_1(\tau T) }  - x_1 \right]\\
\frac{d x_2}{d\tau} & =  \varepsilon \left[ k_2^0 + \frac{k_2}{1+ x_1^2 \cdot w_2(\tau T) }  - x_2 \right]
\end{split}
\end{equation}
with $\varepsilon=T g^\mathrm{p}$. The vector field in \eqref{eq:sys_orig} is time-varying in $\tau$ with period $1$, and it is now in a form amenable for periodic averaging analysis (see Supplementary Material). 

In particular, the average vector field, say $f_\mathrm{av}(x)$, can be obtained by integrating the vector field in \eqref{eq:sys_orig} over a period, yielding
\begin{equation*}
\begin{split}
f_{\mathrm{av},1}(x)&=  \frac{1}{1}\int_0^1 \left( k_1^0 + \frac{k_1}{1+ x_2^2 \cdot w_1(\tau T) }  - x_1 \right) d\tau\\
& = k_1^0 + k_1 \!\! \left( \! \int_0^D \!\!\!\!\! \frac{1}{1+x_2^2 \! \cdot \! 1} d\tau \! + \!\! \int_D^1 \!\! \frac{1}{1+x_2^2 \! \cdot \! \bar{w}_1} d\tau \!\! \right) \! - \! x_1\\
& = k_1^0+ k_1 \left( \frac{D}{1+x_2^2}+ \frac{1-D}{1+x_2^2 \!\cdot \! \bar{w}_1}  \right) -x_1,
\end{split}
\end{equation*}
where we used \eqref{eq:atc_squarewave}, and similarly for $f_{\mathrm{av},2}(x)$,
\begin{equation*}
\begin{split}
f_{\mathrm{av},2}(x)&=  \frac{1}{1}\int_0^1 \left( k_2^0 + \frac{k_2}{1+ x_1^2 \cdot w_2(\tau T) }  - x_2 \right) d\tau\\
& = k_2^0 + k_2 \!\! \left( \! \int_0^D \!\!\!\!\! \frac{1}{1+x_1^2 \!\cdot\! \bar{w}_2} d\tau \! + \!\! \int_D^1 \!\! \frac{1}{1+x_1^2 \! \cdot \! 1} d\tau \!\! \right) \! - \! x_2\\
& = k_2^0+ k_2 \left( \frac{D}{1+x_1^2 \!\cdot\! \bar{w}_2}+ \frac{1-D}{1+x_1^2}  \right) -x_2,
\end{split}
\end{equation*}
where we used \eqref{eq:iptg_squarewave}.

Hence, the resulting \emph{average system} is
\begin{equation}
\label{eq:sys_average}
\begin{split}
\frac{d x_1}{d\tau} & =  \varepsilon \left[k_1^0+ k_1 \left( \frac{D}{1+x_2^2}+ \frac{1-D}{1+x_2^2\cdot \bar{w}_1}  \right) -x_1 \right]\\
\frac{d x_2}{d\tau} & =  \varepsilon \left[ k_2^0+ k_2 \left( \frac{D}{1+x_1^2 \cdot \bar{w}_2}+ \frac{1-D}{1+x_1^2}  \right) -x_2 \right]
\end{split}
\end{equation}

Let $x(\tau,\varepsilon)$ and $x_\mathrm{av}(\varepsilon\tau)$ denote the solutions to \eqref{eq:sys_orig} and \eqref{eq:sys_average}, respectively. Assume $\bar{x}_\mathrm{av}$ is an exponentially stable equilibrium point of the average system \eqref{eq:sys_average}. Let $\Omega$ be a compact subset of its basin of attraction, and assume $x_\mathrm{av}(0)\in\Omega$, and $x(0,\varepsilon)-x_\mathrm{av}(0)=O(\varepsilon)$. Then, from \cite[Theorem 10.4]{khalil2002nonlinear}, there exists a positive parameter $\varepsilon^\ast=T^\ast g^\mathrm{p}$ such that for all $0<\varepsilon<\varepsilon^\ast$ 
\begin{equation}
\label{eq:averaging_bound}
x(\tau,\varepsilon)-x_\mathrm{av}(\varepsilon\tau)=O(\varepsilon)
\end{equation}
for all $\tau>0$. 
That is, solutions $x(\tau,\varepsilon)$ to system \eqref{eq:sys_orig} can be approximated by solutions $x_\mathrm{av}(\varepsilon\tau)$ to \eqref{eq:sys_average} with an error that is proportional to $\varepsilon$. As a consequence, if $\bar{x}_\mathrm{av}$ is the unique equilibrium point of system \eqref{eq:sys_average}, then for all $0<\varepsilon<\varepsilon^\ast$ system \eqref{eq:sys_orig} has a unique, exponentially stable, periodic solution $\bar{x}(\tau,\varepsilon)$ in a $O(\varepsilon)$-neighborhood of $\bar{x}_\mathrm{av}$.
\begin{figure}[!t]
\begin{center}
\includegraphics[width=0.7\linewidth]{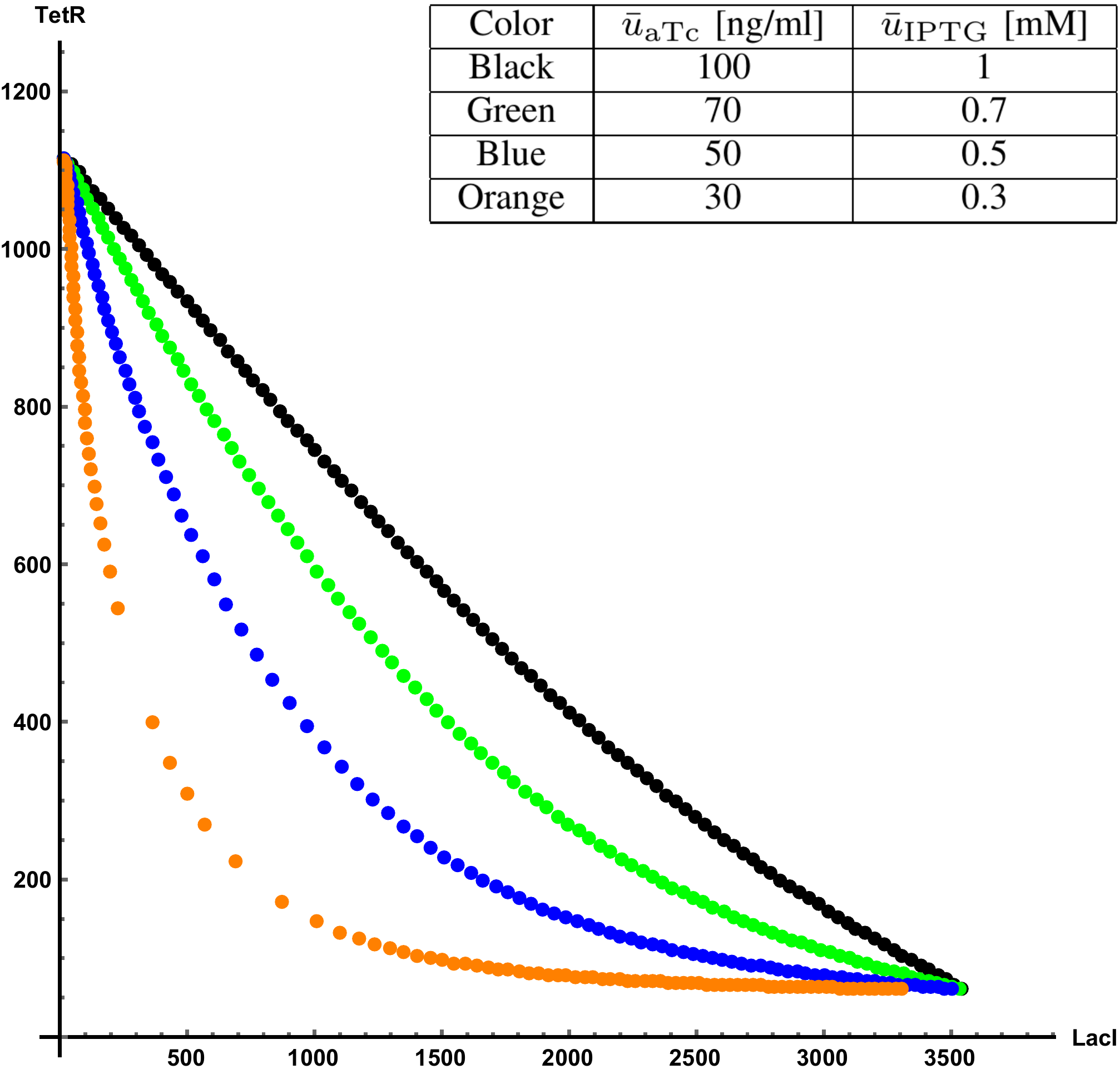}
\caption{Equilibrium points $\bar{x}_\mathrm{av}$ of \eqref{eq:sys_average} as a function of duty cycle $D$ rescaled in arbitrary fluorescence units using \eqref{eq:adim_variables}. Each dot represents the location of the unique stable equilibrium point of system \eqref{eq:sys_average} evaluated for $D$ taking values in the interval $[0,1]$ with increments of $0.01$. }
\label{fig:equilibria_varying_equally}
\end{center}
\end{figure}

The number and position in state space of the equilibrium points $\bar{x}_\mathrm{av}$ of the average system \eqref{eq:sys_average} depend on the specific choice of the amplitudes $\bar{u}_{\mathrm{aTc}}$ and $\bar{u}_{\mathrm{IPTG}}$ of  the pulse waves, and also on the value of the duty cycle $D$.
For example, for the reference values $\bar{u}_{\mathrm{aTc}}=50\, \mathrm{ng/ml}$ and $\bar{u}_{\mathrm{IPTG}}=0.5\, \mathrm{mM}$, system \eqref{eq:sys_average} is monostable and the position of the equilibrium point $\bar{x}_\mathrm{av}$ varies monotonically with $D$ as reported in Figure \ref{fig:equilibria_varying_equally} (blue dots). 
Hence, given certain values of $\bar{u}_{\mathrm{aTc}}$ and $\bar{u}_{\mathrm{IPTG}}$, it is possible to move the position of $\bar{x}_\mathrm{av}$ on the corresponding curve by varying $D$ (Supplementary Figure S1).

The phase portrait of the average system \eqref{eq:sys_average} together with a representative solution of the time-varying system \eqref{eq:sys_orig} for $D$ equal to $0.5$ are depicted in Figure \ref{fig:no_approx_dc_05}, while for $D$ equal to $0.2$ and $0.8$ are reported in Supplementary Figure S2.
The parameter $\varepsilon$ has been set to $0.1$ which corresponds to a forcing period $T=\varepsilon/g^\mathrm{p}\approx 6\, \mathrm{min}$, and the system has been simulated for $t_f=\tau_f\, T\approx 50 \cdot 6= 300 \, \mathrm{min}$.
Larger values of $\varepsilon$ correspond to larger values of the forcing period $T$. In turn, from \eqref{eq:averaging_bound}, this also implies that the solution $x(\tau,\varepsilon)$ of \eqref{eq:sys_orig} will asymptotically converge to a periodic solution $\bar{x}(\tau,\varepsilon)$ contained in a larger set (Figure \ref{fig:no_approx_dc_05_eps_10}), and hence to a worse approximation (see also Supplementary Figure S6 for their time evolution).
\begin{figure}
\centering
\subfigure[$D=0.5$, $T\approx 6\,\mathrm{min}$  ($\varepsilon=0.1$).]
{
\includegraphics[width=0.7\linewidth]{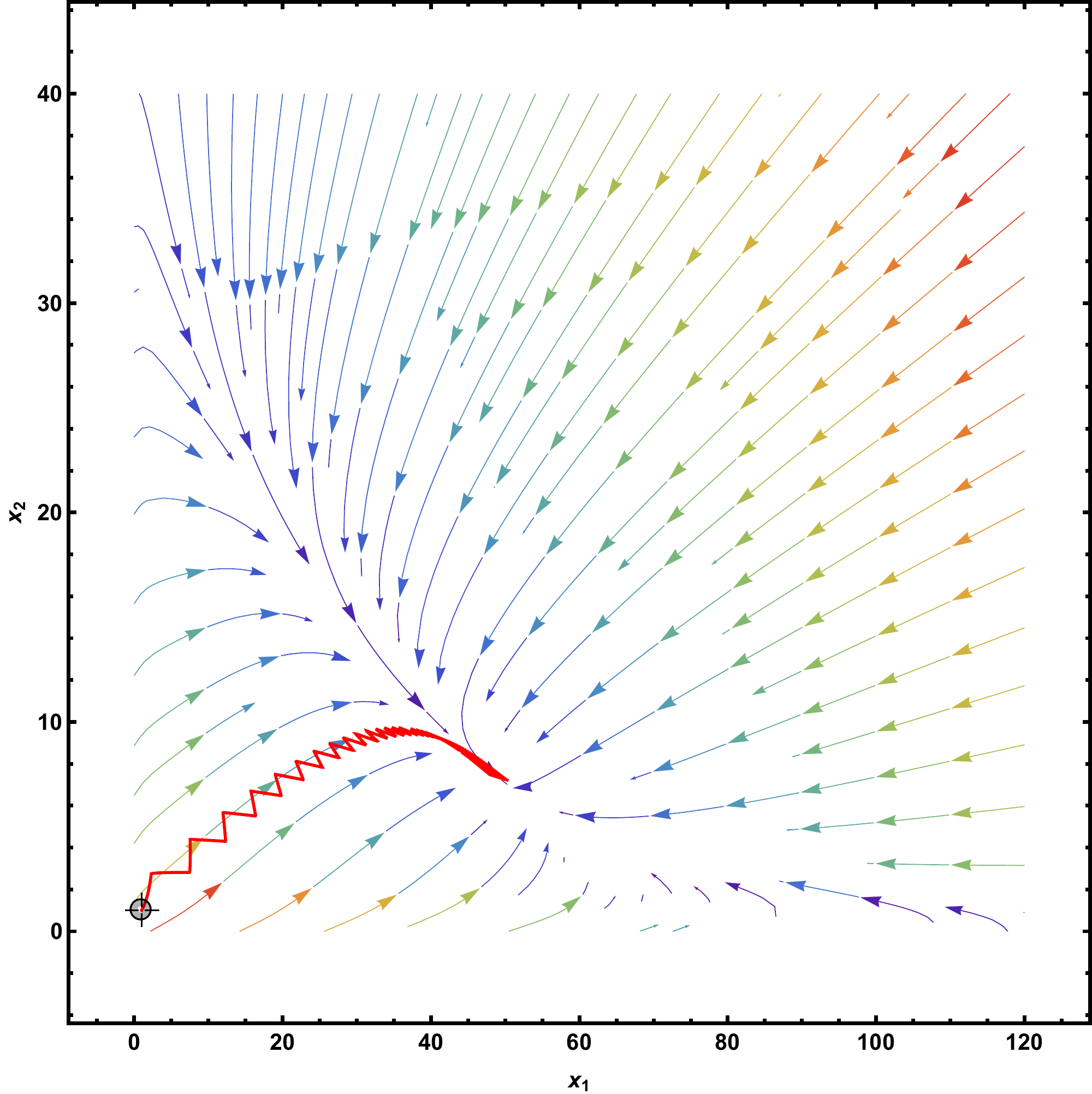}
\label{fig:no_approx_dc_05}
}
\\
\subfigure[$D=0.5$, $T\approx 180\,\mathrm{min}$ ($\varepsilon=3$).]
{
\includegraphics[width=0.7\linewidth]{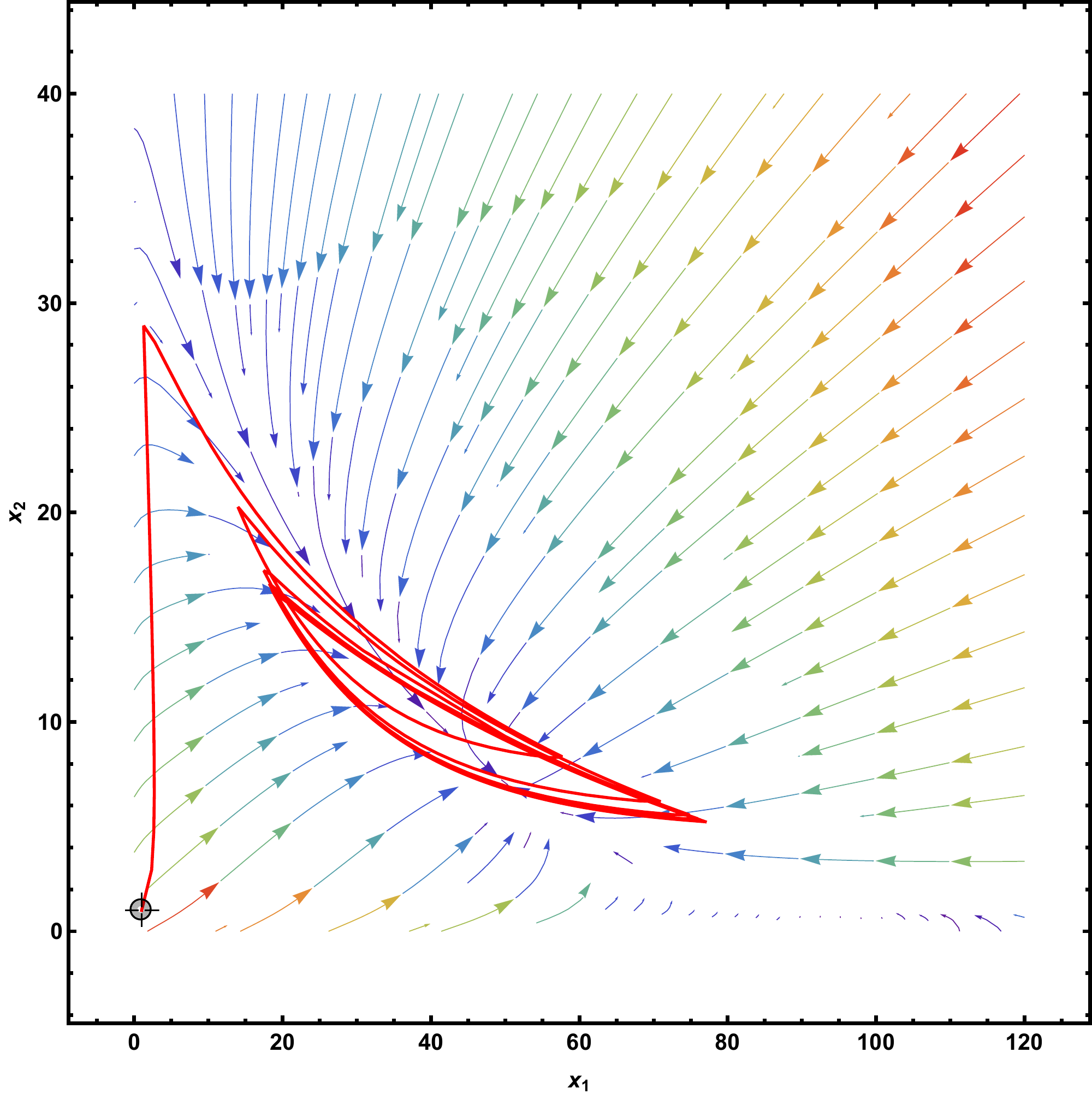}
\label{fig:no_approx_dc_05_eps_10}
}
\caption{Background: phase portrait of the average system \eqref{eq:sys_average}. Red line: the solution of the time-varying system \eqref{eq:sys_orig} with $\bar{u}_{\mathrm{aTc}}=50\, \mathrm{ng/ml}$ and $\bar{u}_{\mathrm{IPTG}}=0.5\, \mathrm{mM}$ from initial condition ${[1,1]}^{\mathsf{T}}$.}
\label{fig:pplane_2}
\end{figure}
%
%
%%%%%%%%%%%%%%%%%%%%%%%%%%%%%%%%%%%%%%%%%%%%%%%%%%%%%%%%%%%
%
%
\section{Diffusion effects}
\label{sec:simulations}
The analysis in the previous section was conducted under Assumption \ref{ass:diffusion}. As already mentioned before, the cell membrane acts as a low-pass filter, hence, when Assumption \ref{ass:diffusion} is dropped, $aTc(t)$ and $IPTG(t)$ will not anymore be ideal pulse waves but their filtered versions through the cell membrane. Therefore, in order for the average system \eqref{eq:sys_average} to continue being a good approximation of the actual cell response, the cut-off frequency of the two low-pass filters should be sufficiently higher than the fundamental frequency $1/T$ of the input pulse waves. However, due to the inevitable attenuation of high-frequency harmonics, there will always be a mismatch between the actual mean response of the cell and the value predicted by \eqref{eq:sys_average}. 

\begin{figure}[!t]
\begin{center}
\includegraphics[width=0.7\linewidth]{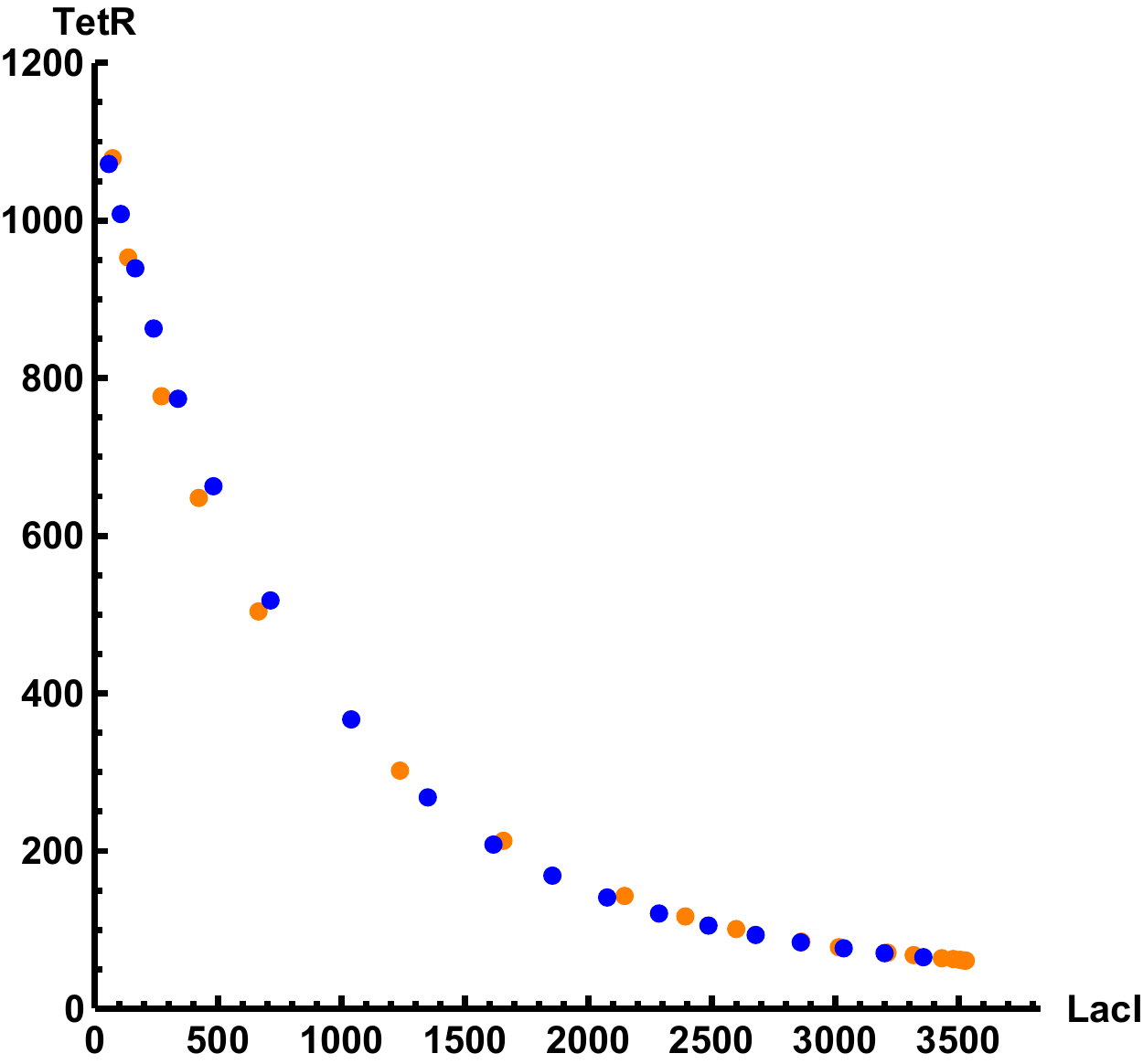}
\caption{Orange dots: Mean-value, evaluated at regime, of the response of system \eqref{eq:transcr_laci}-\eqref{eq:transl_tetr} (with membrane dynamics \eqref{eq:diffusion_atc}-\eqref{eq:diffusion_iptg}) to PWM inputs with $T=240\,\mathrm{min}$ and varying $D$ from $0.05$ to $0.95$ with increments of $0.05$. Blue dots: corresponding equilibrium point $\bar{x}_{\mathrm{av}}(D)$ of system \eqref{eq:sys_average} rescaled in a.u. using \eqref{eq:adim_variables}. Amplitude of pulse waves set to $\bar{u}_{\mathrm{aTc}}=50\, \mathrm{ng/ml}$ and $\bar{u}_{\mathrm{IPTG}}=0.5\, \mathrm{mM}$.}
\label{fig:sim50}
\end{center}
\end{figure}

The effects of relaxing Assumption \ref{ass:diffusion} on the time response of system \eqref{eq:sys_orig} can be observed in Supplementary Figures S4-S5.
%In Figure \ref{fig:sim50} we report the comparison, for one choice of the amplitudes of the PWM signals (see Supplementary Figure S3 for a different choice) and varying the duty cycle $D$, between 
The mean steady-state response of the complete four-dimensional system \eqref{eq:transcr_laci}-\eqref{eq:transl_tetr} with diffusion dynamics \eqref{eq:diffusion_atc}-\eqref{eq:diffusion_iptg} is compared in Figure \ref{fig:sim50}, and the corresponding equilibrium point $\bar{x}_{\mathrm{av}}(D)$ predicted by the autonomous two-dimensional average system \eqref{eq:sys_average}, for a representative value of the PWM amplitudes and different values of $D$ (see Supplementary Figure S3 for a different choice).
Although as expected there is no perfect matching between the two, the observed behavior is well captured by the average system. Note that in regulation problems, this mismatch can be compensated by designing an adequate feedback action.

When, on the other hand, the cut-off frequency of one of the filters is lower than the frequency $1/T$ of the input pulse waves, the input signal will be highly attenuated, resulting in the simple regulation of the toggle switch to either one of the stable equilibrium points (a phenomenon that was reported in the experiments described in \cite[Supplementary Figure 8]{lugagne2017balancing}). A similar phenomenon can also occur when the duty cycle is close to $0$ or $1$. Indeed, close to these values, the amplitude of the harmonics of the pulse wave is $|a_n|=\left| \frac{2 \bar{u}}{n\,\pi} \sin(n\pi D)\right| \approx 2\bar{u}D$, therefore low-frequency harmonics will have amplitudes similar to those of high-frequency ones, and the pulse wave will be highly attenuated.

%
%
%%%%%%%%%%%%%%%%%%%%%%%%%%%%%%%%%%%%%%%%%%%%%%%%%%%%%%%%%%%%%
%
%
\section{Perspectives for control}
\label{sec:control}
\begin{figure}[!t]
\begin{center}
\includegraphics[width=\columnwidth]{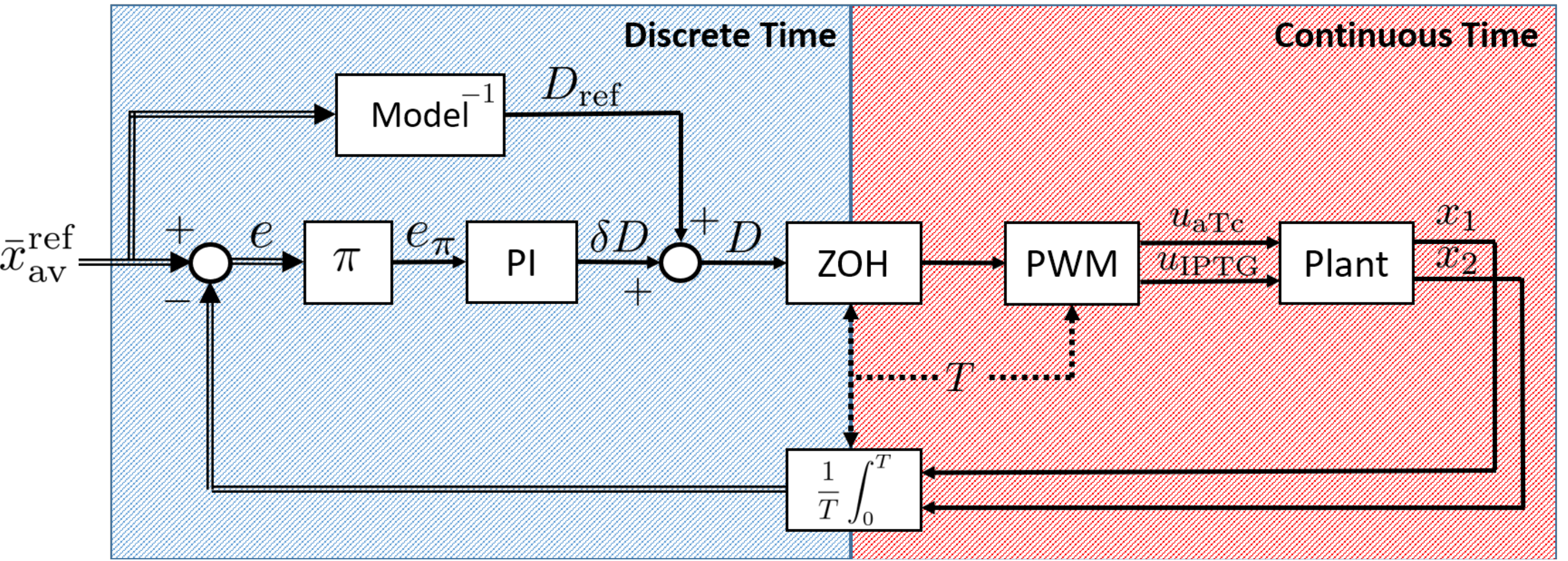}
\caption{External controller for the regulation of the mean-response of a genetic toggle switch.}
\label{fig:block_controller}
\end{center}
\end{figure}

We wish to emphasize that the analytical results derived here can be exploited for the synthesis of external controllers to regulate the mean-value of the output response of the genetic toggle switch. Specifically, we propose the control schematic shown in Figure \ref{fig:block_controller} which is currently under development and will be presented elsewhere. Indeed, as done in Figure \ref{fig:equilibria_varying_equally}, it is possible to numerically compute $\bar{x}_\mathrm{av}$ as a function of $\bar{u}_\mathrm{aTc}$, $\bar{u}_\mathrm{IPTG}$ and $D$, and get interpolating curves $\Gamma_{\bar{u}_\mathrm{aTc},\bar{u}_\mathrm{IPTG}}(D)$. From these one can then obtain, for given values of the amplitude $\bar{u}_\mathrm{aTc}$ and $\bar{u}_\mathrm{IPTG}$, the duty cycle $D_\mathrm{ref}$ corresponding to the desired average set-point $\bar{x}_\mathrm{av}^\mathrm{ref}$, that is $D_{\mathrm{ref}}=\Gamma_{\bar{u}_\mathrm{aTc},\bar{u}_\mathrm{IPTG}}^{-1}(\bar{x}_\mathrm{av}^\mathrm{ref})$. The mismatch $e$ between the measured mean-value of the plant outputs and $\bar{x}_\mathrm{av}^\mathrm{ref}$ is then projected by $\pi$ onto the curve $\Gamma_{\bar{u}_\mathrm{aTc},\bar{u}_\mathrm{IPTG}}$ and compensated by a PI controller. The control scheme should also take into account the effects of the sampling time and of the slow transients.
%
%

%\addtolength{\textheight}{-9cm}   % This command serves to balance the column lengths
\section{Conclusions}
\label{sec:conclusions}
We derived and analyzed a model to capture the response of the genetic toggle switch to mutually exclusive PWM inputs observed experimentally in \cite{lugagne2017balancing}. The analysis was based on the assumption that the diffusion of inducer molecules across the cell membrane is instantaneous. From this, using the periodic averaging method for nonlinear systems, we derived an autonomous vector field that describes the dynamics of the mean-value of the periodic solutions of the original system. After discussing the predictions of the model under the assumption of instantaneous diffusion, we relaxed this assumption so that the input signals become filtered versions of themselves worsening the predictions.
 
However, even if it is not possible to eliminate the attenuation due to the cell membrane, our analysis shows that to mitigate its effects the frequency $1/T$ of the input pulse waves should be chosen sufficiently lower than the cut-off frequency of the low-pass membrane filter, and extreme values of the duty cycle $D$ should be avoided.
At the same time, we find that to avoid large oscillations around $\bar{x}_\mathrm{av}$, the parameter $\varepsilon=T g^\mathrm{p}$, that is the ratio between the time-scales of the forcing inputs and system dynamics, should be taken as small as possible, e.g., for fixed $T$, by cooling down the temperature of the growth medium and thus reducing the cell growth rate and therefore $g^\mathrm{p}$.

Future work will be aimed at quantifying the effects of the attenuation of the input signals due to the cell membrane to improve the predictions of our model, and at implementing and validating (in-silico and in-vivo) external controllers, also capable of modulating the ON/OFF values of the pulse waves. 
Furthermore, we also plan to investigate the effect that different classes of periodic forcing could have on the variance of the response of a population of cells with extrinsic noise.

\footnotesize
\section*{ACKNOWLEDGMENT}

The authors wish to acknowledge support from the research project COSY-BIO (Control Engineering of Biological Systems for Reliable Synthetic Biology Applications) funded by the European Union's Horizon 2020 research and innovation programme under grant agreement No 766840.
%
%
%%%%%%%%%%%%%%%%%%%%%%%%%%%%%%%%%%%%%%%%%%%%%%%%%%%%%%%%%%%%%%%%%%%%%%%%%%%%%%%%
%
%
\bibliographystyle{IEEEtran} 
\bibliography{refs} 
\newpage
\onecolumn

\section*{Supplementary Material}
\subsection*{Periodic averaging}
\label{sec:averaging_method}
We recall here that, from \cite[Theorem 10.4]{khalil2002nonlinear}, the periodic averaging method says that the solutions of the system
\begin{equation} \tag{i}
\label{eq:averaging_sys_original}
\dot{x}=\varepsilon f(t,x,\varepsilon)
\end{equation}
where $f(\cdot)$ is sufficiently smooth with respect to $(x,\varepsilon)$, and $T$-periodic and measurable in $t$, can be approximated by an autonomous average system
\begin{equation} \tag{ii}
\label{eq:averaging_sys_averaged}
\dot{x}=\varepsilon f_\mathrm{av}(x)
\end{equation}
where
$
f_\mathrm{av}(x)=\frac{1}{T}\int_0^T f(s,x,0)ds.
$
More precisely, let $x(t,\varepsilon)$ and $x_\mathrm{av}(\varepsilon t)$ denote the solutions of \eqref{eq:averaging_sys_original} and \eqref{eq:averaging_sys_averaged}, respectively. If system \eqref{eq:averaging_sys_averaged} has an exponentially stable equilibrium point $\bar{x}_\mathrm{av}$, then there exist positive constants $\varepsilon^\ast$ and $k$ such that, for all $0<\varepsilon<\varepsilon^\ast$, system \eqref{eq:averaging_sys_original} has a unique, exponentially stable, $T$-periodic solution $\bar{x}(t,\varepsilon)$ in a $O(\varepsilon)$-neighborhood of $\bar{x}_\mathrm{av}$, that is $\lVert \bar{x}(t,\varepsilon)-\bar{x}_\mathrm{av} \rVert\leq k\varepsilon$. Moreover, if the initial conditions are such that $x(0,\varepsilon)-x_\mathrm{av}(0)=O(\varepsilon)$, then $x(t,\varepsilon)-x_\mathrm{av}(\varepsilon t)=O(\varepsilon)$, for all $t\geq 0$.
\subsection*{Nondimensionalization of system (1)-(4)}
\label{sec:nondimensionalization}
Equations (1) and (2) can be rewritten as 

\footnotesize
\begin{align*}
& \begin{aligned}
\frac{d\, mRNA_{\mathrm{LacI}}}{dt}=\; & \kappa_\mathrm{L}^\mathrm{m0} + \frac{\kappa_\mathrm{L}^\mathrm{m}}{1+ \left( \frac{TetR}{\theta_{\mathrm{TetR}} } \right)^{\eta_{\mathrm{TetR}}} \cdot w_1(t) }   - g_\mathrm{L}^\mathrm{m} \cdot mRNA_{\mathrm{LacI}}
\end{aligned}
\\
& \begin{aligned}
\frac{d\, mRNA_{\mathrm{TetR}}}{dt}=\; & \kappa_\mathrm{T}^\mathrm{m0} + \frac{\kappa_\mathrm{T}^\mathrm{m}}{1+ \left( \frac{LacI}{\theta_{\mathrm{LacI}} }  \right)^{\eta_{\mathrm{LacI}}} \cdot w_2(t) }  - g_\mathrm{T}^\mathrm{m} \cdot mRNA_{\mathrm{TetR}}
\end{aligned}
\end{align*}
\normalsize
where $w_1(t)$ and $w_2(t)$ are defined in (11)-(12).

Now, taking into account that mRNA molecules are degraded faster than other molecules, we can obtain a quasi-steady state approximation by setting $\frac{d\, mRNA_{\mathrm{LacI}}}{dt}=0$ and  $\frac{d\, mRNA_{\mathrm{TetR}}}{dt}=0$, yielding 

\footnotesize
\begin{align*}
mRNA_{\mathrm{LacI}}= \frac{\kappa_\mathrm{L}^\mathrm{m0}}{g_\mathrm{L}^\mathrm{m}} + \frac{\kappa_\mathrm{L}^\mathrm{m}}{g_\mathrm{L}^\mathrm{m}}\,\frac{1}{1+ \left( \frac{TetR}{\theta_{\mathrm{TetR}} } \right)^{\eta_{\mathrm{TetR}}} \cdot w_1(t) } \\
mRNA_{\mathrm{TetR}}= \frac{\kappa_\mathrm{T}^\mathrm{m0}}{g_\mathrm{T}^\mathrm{m}} + \frac{\kappa_\mathrm{T}^\mathrm{m}}{g_\mathrm{T}^\mathrm{m}}\,\frac{1}{1+ \left( \frac{LacI}{\theta_{\mathrm{LacI}} } \right)^{\eta_{\mathrm{LacI}}} \cdot w_2(t) }
\end{align*}
\normalsize
Assuming that LacI and TetR proteins degrade at the same rate, that is $g_\mathrm{L}^\mathrm{p}=g_\mathrm{T}^\mathrm{p}=g^\mathrm{p}$, that $\eta_{\mathrm{LacI}}=\eta_{\mathrm{TetR}}=2$, and using the dimensionless state variables and time $t'=g^\mathrm{p}\, t$, $x_1=\frac{LacI}{\theta_{\mathrm{LacI}} }$, $x_2=\frac{TetR}{\theta_{\mathrm{TetR}}}$, we obtain from (3)

\footnotesize
\begin{equation*}
\begin{split}
\frac{dx_1}{dt'} =  \frac{\kappa_\mathrm{L}^\mathrm{p}}{\theta_{\mathrm{LacI}}\, g^\mathrm{p} } \, mRNA_{\mathrm{LacI}} - x_1
 = \frac{\kappa_\mathrm{L}^\mathrm{p}}{\theta_{\mathrm{LacI}}\, g^\mathrm{p} } \, \left[ \frac{\kappa_\mathrm{L}^\mathrm{m0}}{g_\mathrm{L}^\mathrm{m}} + \frac{\kappa_\mathrm{L}^\mathrm{m}}{g_\mathrm{L}^\mathrm{m}}\,\frac{1}{1+ x_2^2 \cdot w_1(t'/g^\mathrm{p}) } \right] - x_1
 = k_1^0 + \frac{k_1}{1+ x_2^2 \cdot w_1(t'/g^\mathrm{p}) }  - x_1
\end{split}
\end{equation*}
\normalsize
with $k_1^0$ and $k_1$ as in (9), and from (4)

\footnotesize
\begin{equation*}
\begin{split}
\frac{dx_2}{dt'} =  \frac{\kappa_\mathrm{T}^\mathrm{p}}{\theta_{\mathrm{TetR}}\, g^\mathrm{p} } \, mRNA_{\mathrm{TetR}} - x_1
 = \frac{\kappa_\mathrm{T}^\mathrm{p}}{\theta_{\mathrm{TetR}}\, g^\mathrm{p} } \, \left[ \frac{\kappa_\mathrm{T}^\mathrm{m0}}{g_\mathrm{T}^\mathrm{m}} + \frac{\kappa_\mathrm{T}^\mathrm{m}}{g_\mathrm{T}^\mathrm{m}}\,\frac{1}{1+ x_1^2 \cdot w_2(t'/g^\mathrm{p}) } \right] - x_2
 = k_2^0 + \frac{k_2}{1+ x_1^2 \cdot w_2(t'/g^\mathrm{p}) }  - x_2
\end{split}
\end{equation*}
\normalsize
with $k_2^0$, $k_2$ as in (10).

\clearpage
\begin{figure}[h]
\centering
\subfigure[Equilibrium points for $\bar{u}_{\mathrm{aTc}}=100\, \mathrm{ng/ml}$ and different values of $\bar{u}_{\mathrm{IPTG}}$.]
{
\includegraphics[width=0.4\linewidth]{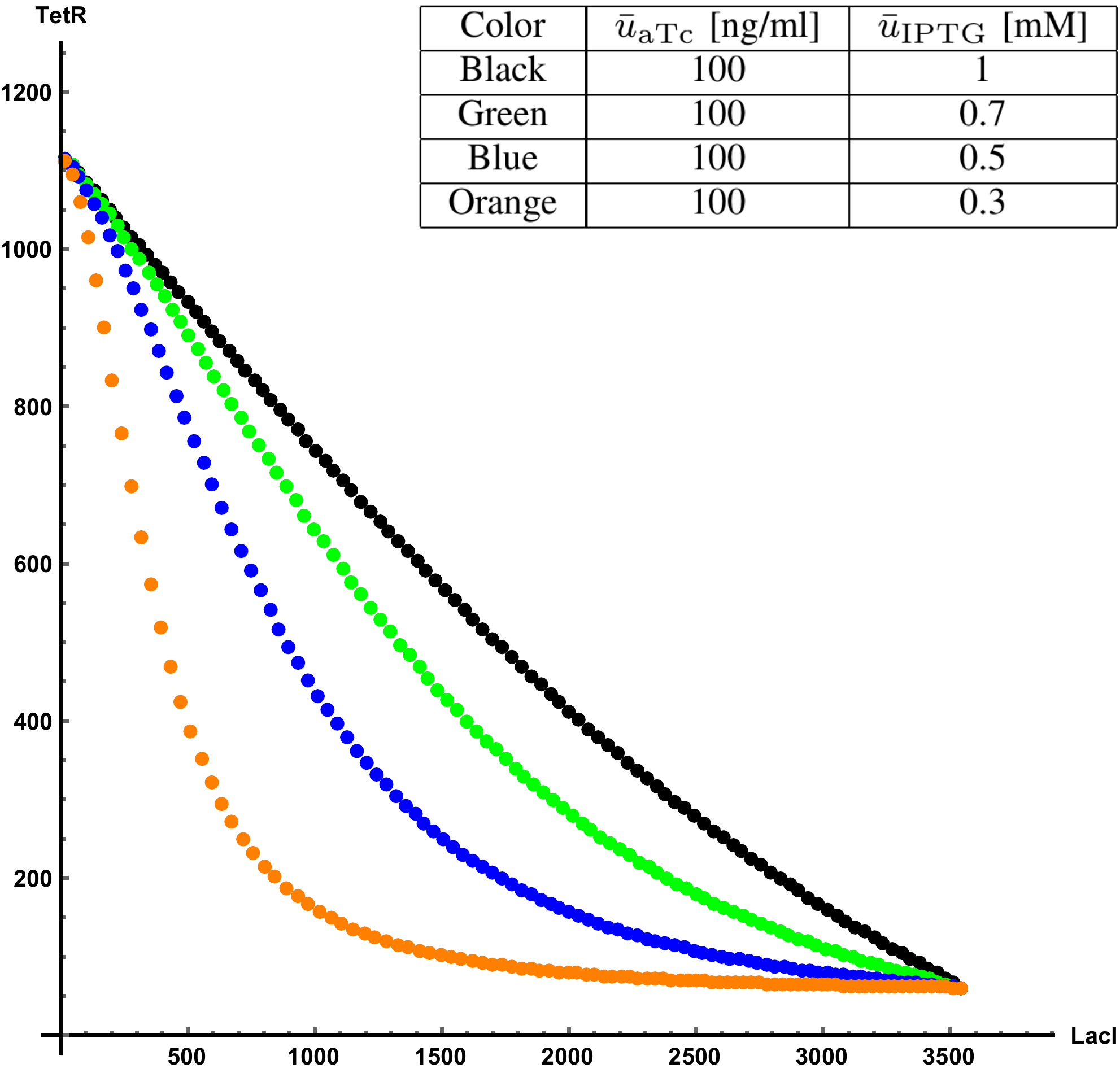}
\label{fig:equilibria_high_atc_varying_iptg}
}
\subfigure[Equilibrium points for $\bar{u}_{\mathrm{IPTG}}=1\, \mathrm{mM}$ and different values of $\bar{u}_{\mathrm{aTc}}$.]
{
\includegraphics[width=0.4\linewidth]{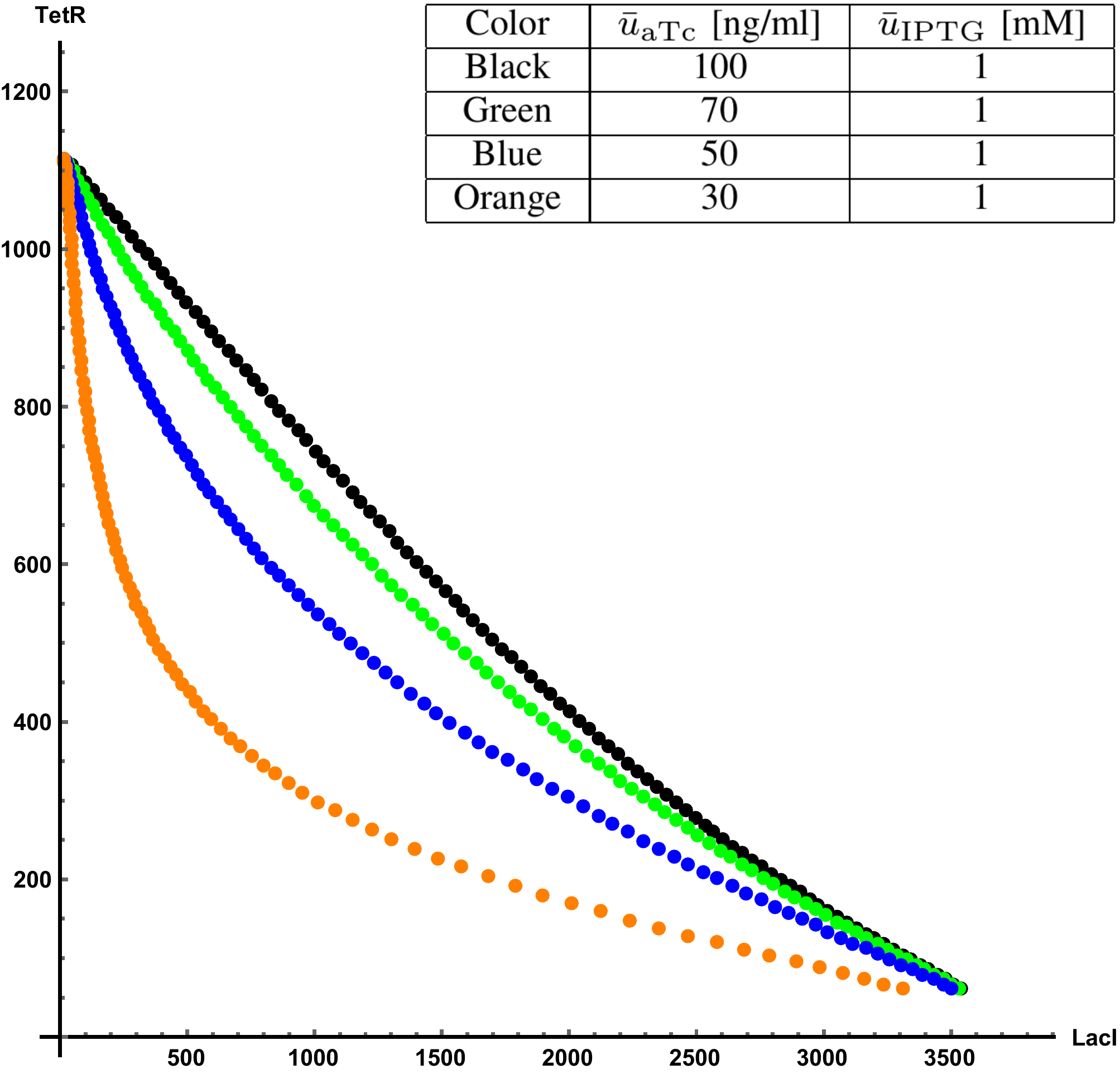}
\label{fig:equilibria_high_iptg_varying_atc}
}
\caption{Equilibrium points $\bar{x}_\mathrm{av}$ of (20) as a function of duty cycle $D$ rescaled in arbitrary fluorescence units using (7). Each dot represents the location of the unique stable equilibrium point of system (20) evaluated for $D$ taking values in the interval $[0,1]$ with increments of $0.01$. }
\label{fig:equilibria}
\end{figure}
\begin{figure}[h]
\centering
\subfigure[$D=0.2$, $T\approx 6\,\mathrm{min}$  ($\varepsilon=0.1$).]
{
\includegraphics[width=0.4\linewidth]{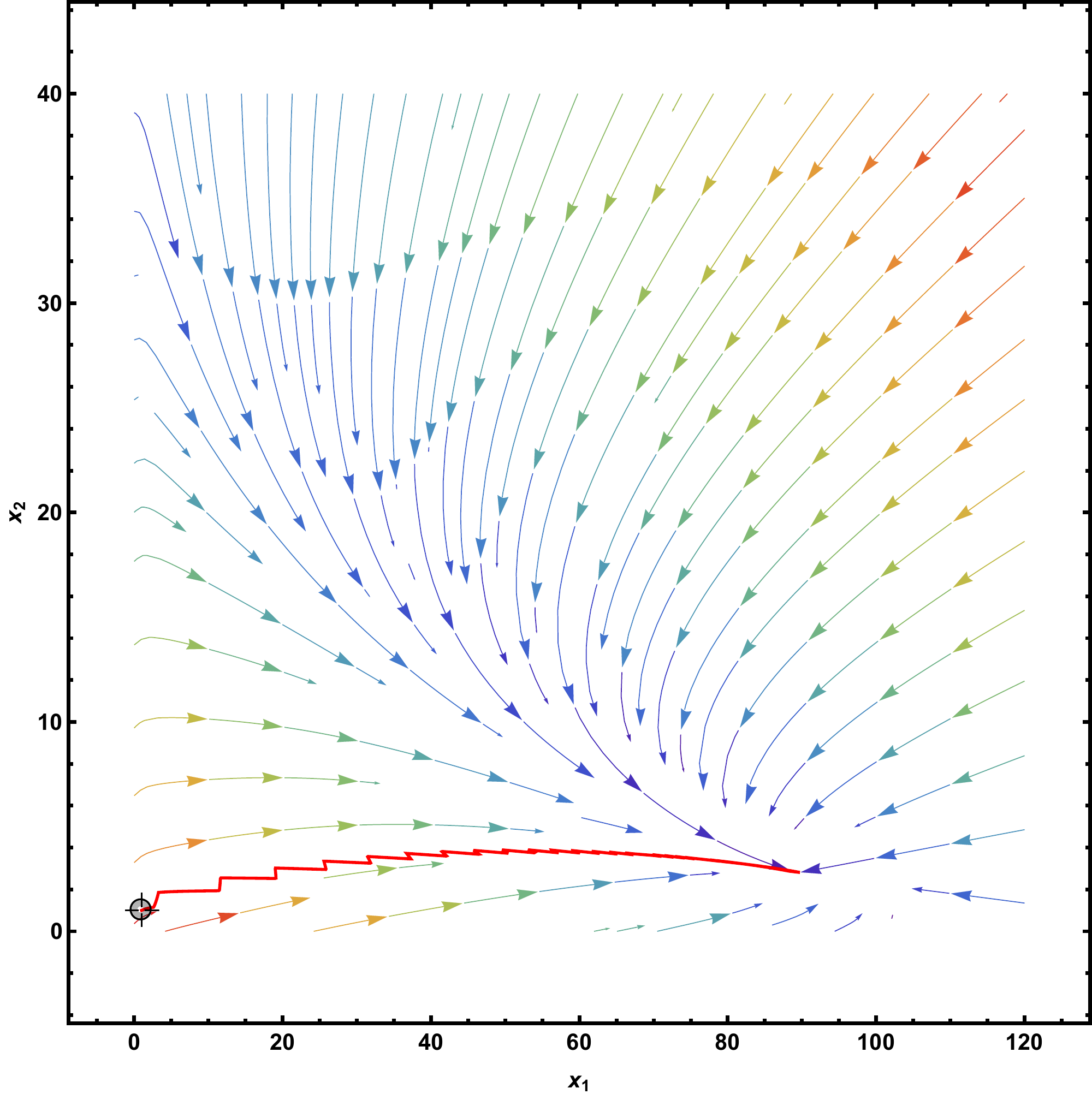}
\label{fig:no_approx_dc_02}
}
\subfigure[$D=0.8$, $T\approx 6\,\mathrm{min}$  ($\varepsilon=0.1$).]
{
\includegraphics[width=0.4\linewidth]{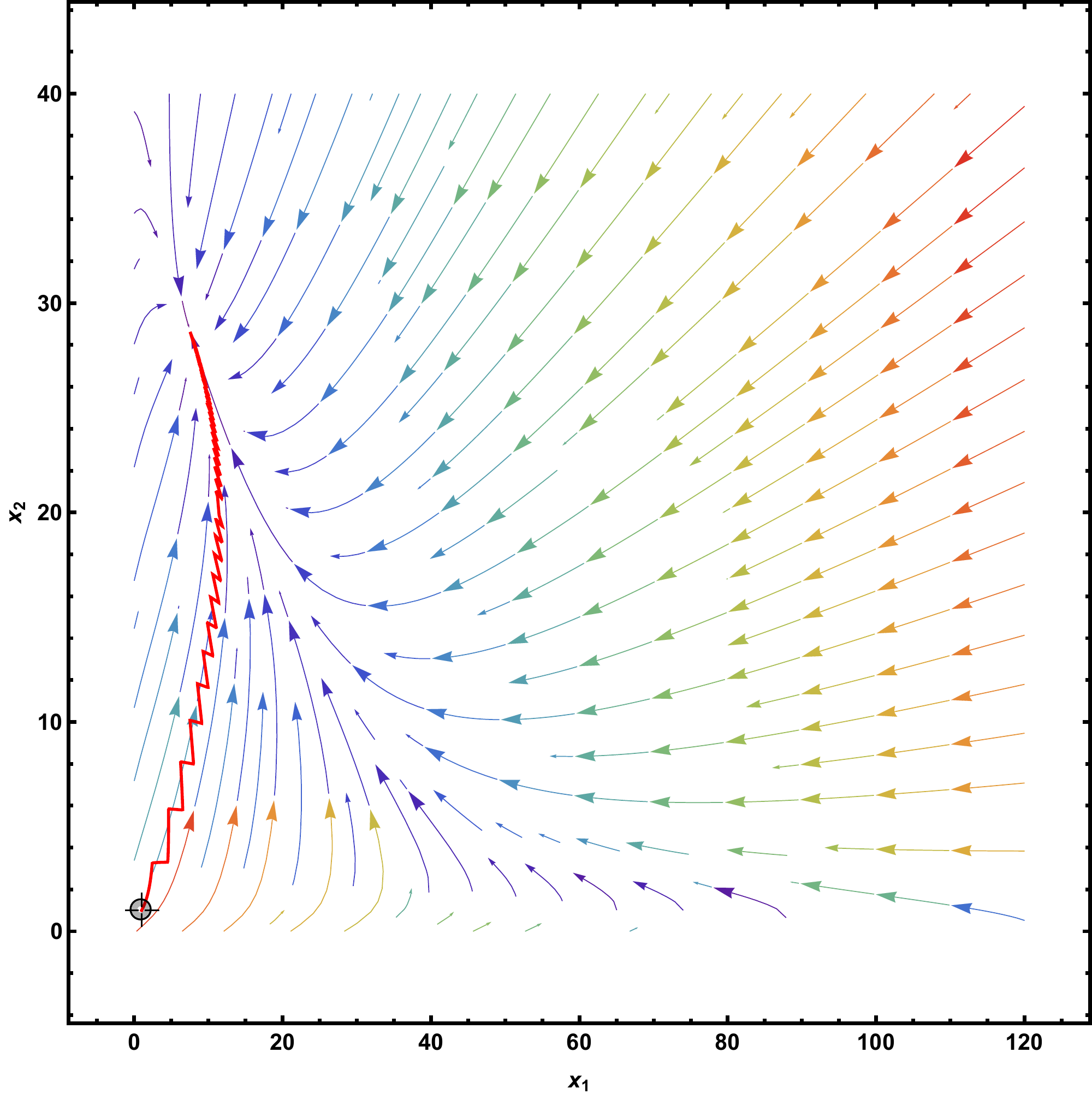}
\label{fig:no_approx_dc_08}
}
\caption{Background: phase portrait of the average system (20). Red line: the solution of the time-varying system (19) with $\bar{u}_{\mathrm{aTc}}=50\, \mathrm{ng/ml}$ and $\bar{u}_{\mathrm{IPTG}}=0.5\, \mathrm{mM}$ from initial condition ${[1,1]}^{\mathsf{T}}$.}
\label{fig:pplane_1}
\end{figure}
\clearpage
\begin{figure}[!h]
\begin{center}
\includegraphics[width=0.4\linewidth]{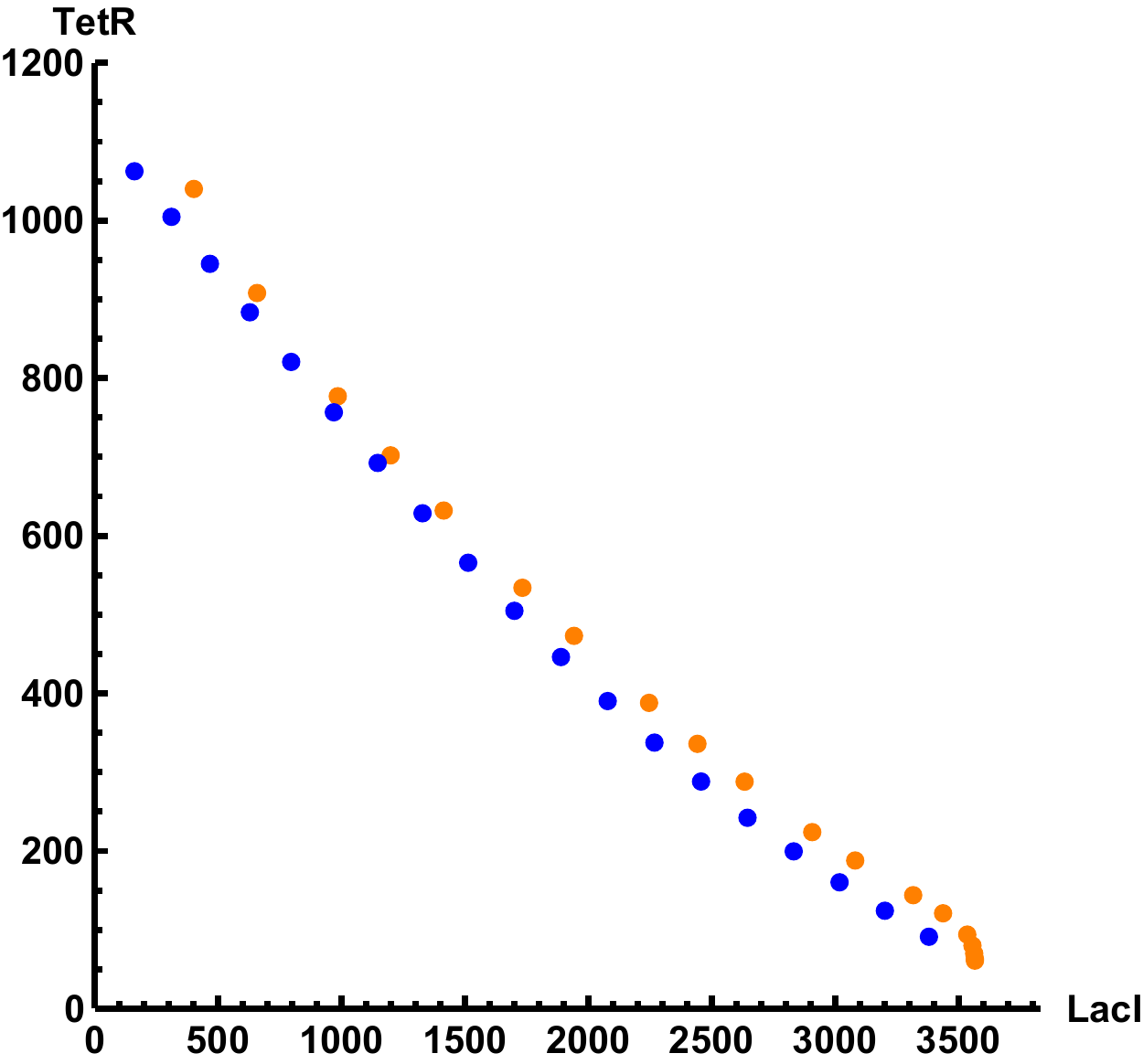}
\caption{Orange dots: Mean-value, evaluated at regime, of the response of system (1)-(4) (with membrane dynamics (5)-(6)) to PWM inputs with $T=240\,\mathrm{min}$ and varying $D$ from $0.05$ to $0.95$ with increments of $0.05$. Blue dots: corresponding equilibrium point $\bar{x}_{\mathrm{av}}(D)$ of system (20). Amplitude of pulse waves set to $\bar{u}_{\mathrm{aTc}}=100\, \mathrm{ng/ml}$ and $\bar{u}_{\mathrm{IPTG}}=1\, \mathrm{mM}$.}
\label{fig:sim10}
\end{center}
\end{figure}

%%%%%%%%%%%%%%%%%%%%%%%%%%%%%%%%%%%%%%%%%%%%%%%%%%%%%%%%%%%%%%%%%%%%%%%%%%%%%

\begin{figure}[!h]
\centering
\subfigure[With instantaneous diffusion]
{
\includegraphics[width=0.4\linewidth]{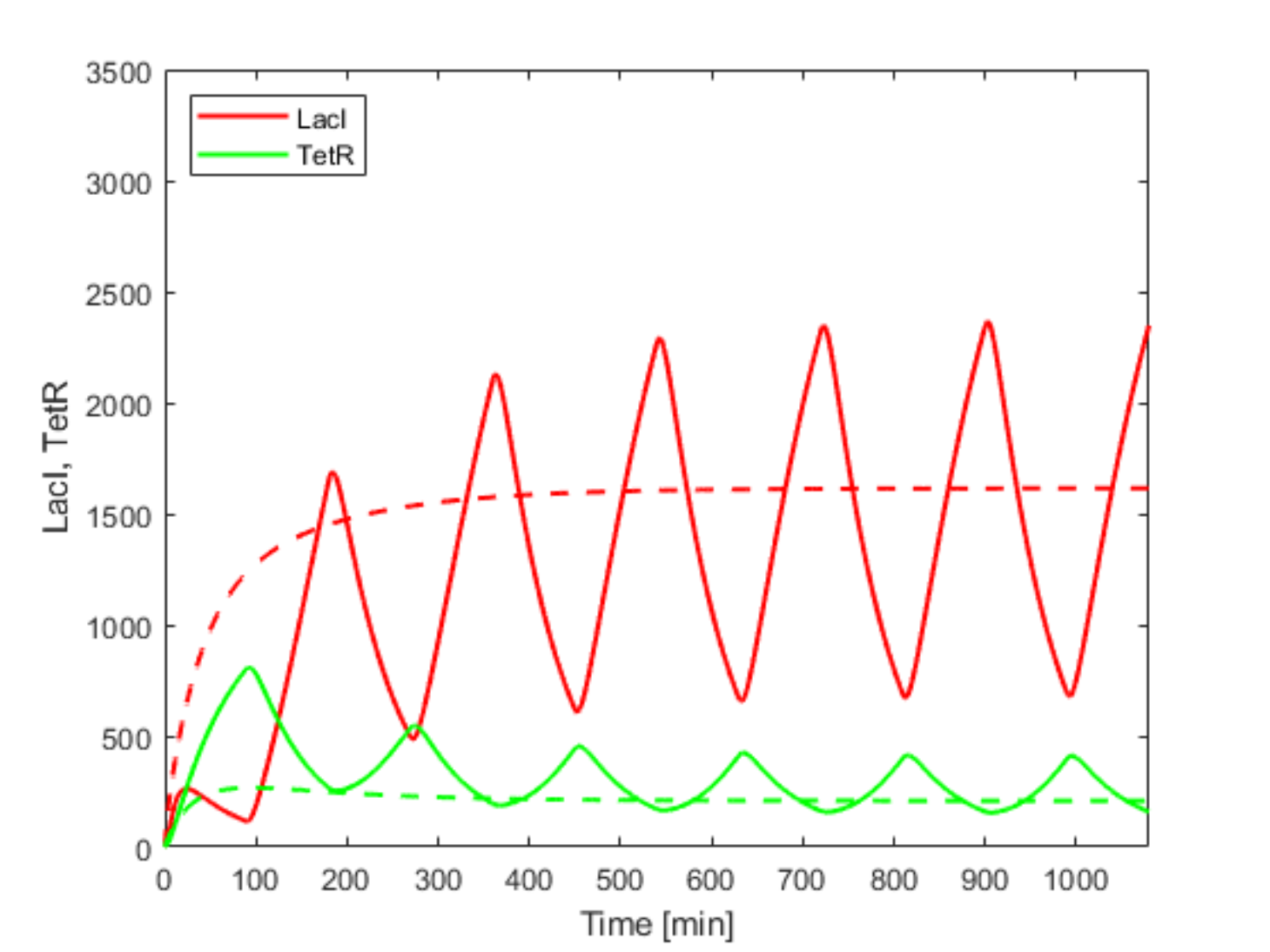}
\label{fig:tplot_atc050_iptg050_d050_no_diff}
}
\subfigure[With diffusion dynamics (5)-(6)]
{
\includegraphics[width=0.4\linewidth]{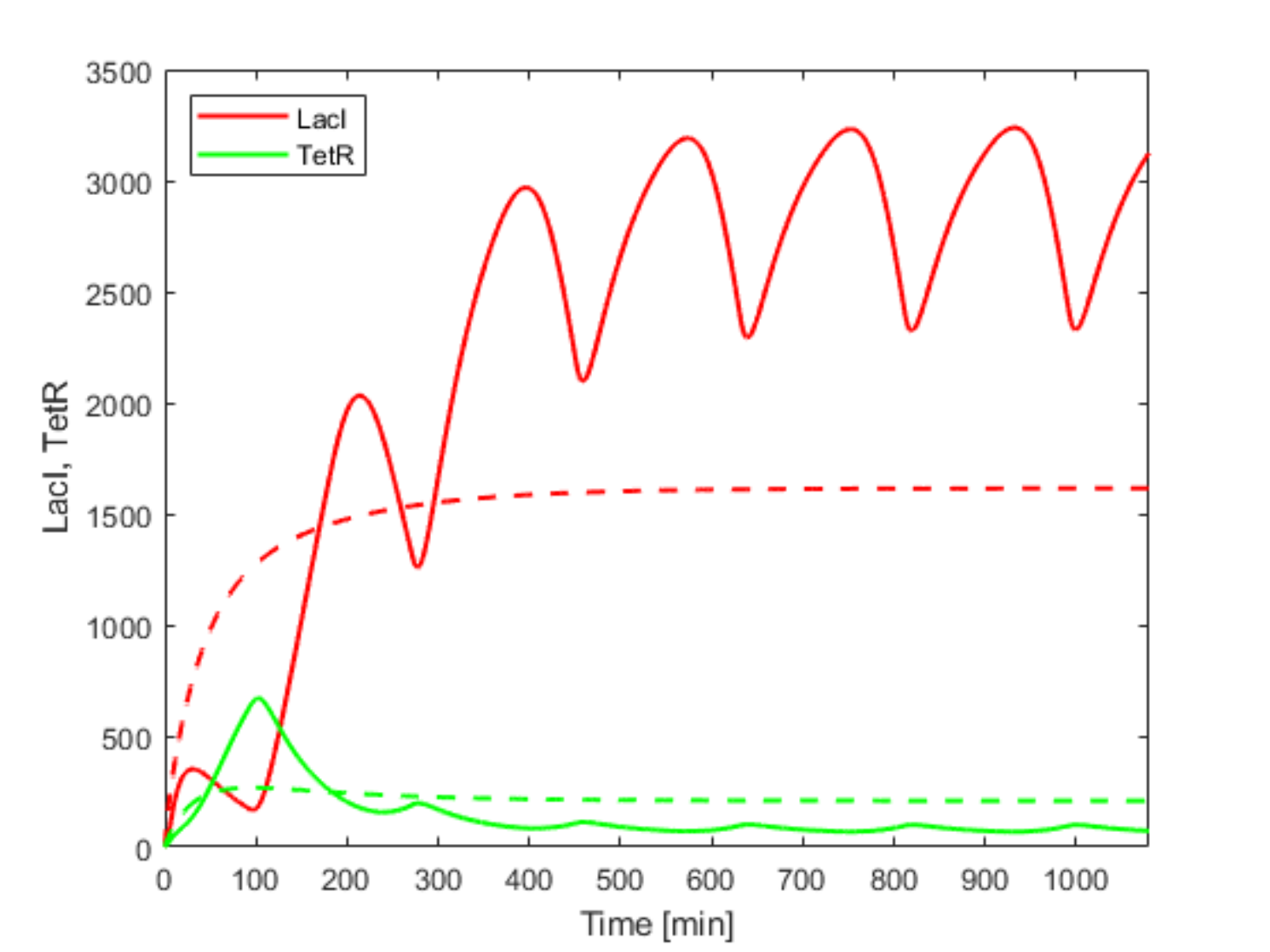}
\label{fig:tplot_atc050_iptg050_d050_diff}
}
\caption{Time evolution of the time-varying system (19) (in solid lines) and of the average system (20) (dashed lines) from initial conditions $[1,\, 1]^\mathsf{T}$ with $\bar{u}_\mathrm{aTc}=50\, \mathrm{ng/ml}$, $\bar{u}_{\mathrm{IPTG}}=0.5\, \mathrm{mM}$, $T=180\,\mathrm{min}$, $D=0.5$.}
\label{fig:tplot_d050}
\end{figure}

\begin{figure}[!h]
\centering
\subfigure[With instantaneous diffusion]
{
\includegraphics[width=0.4\linewidth]{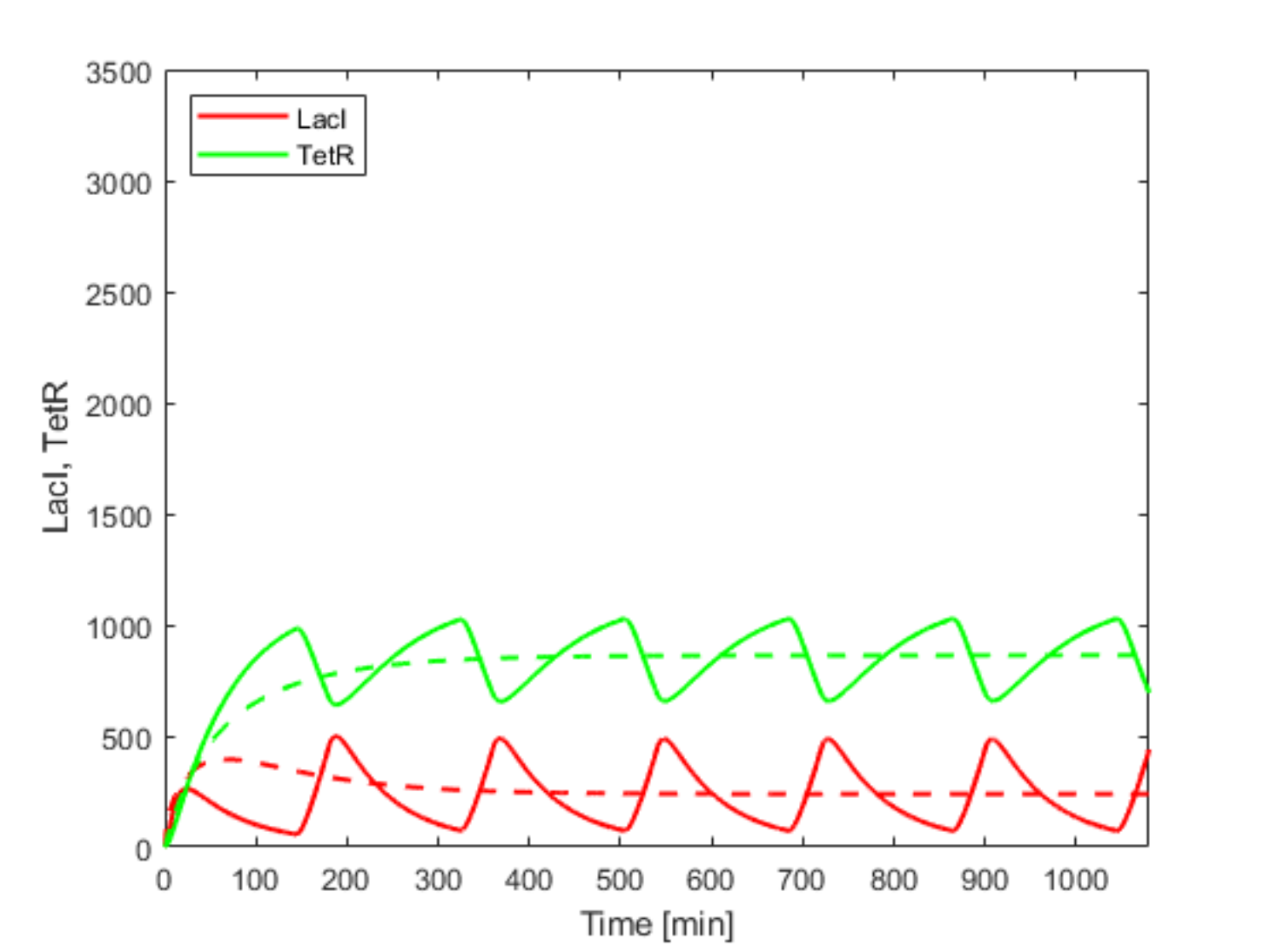}
\label{fig:tplot_atc050_iptg050_d080_no_diff}
}
\subfigure[With diffusion dynamics (5)-(6)]
{
\includegraphics[width=0.4\linewidth]{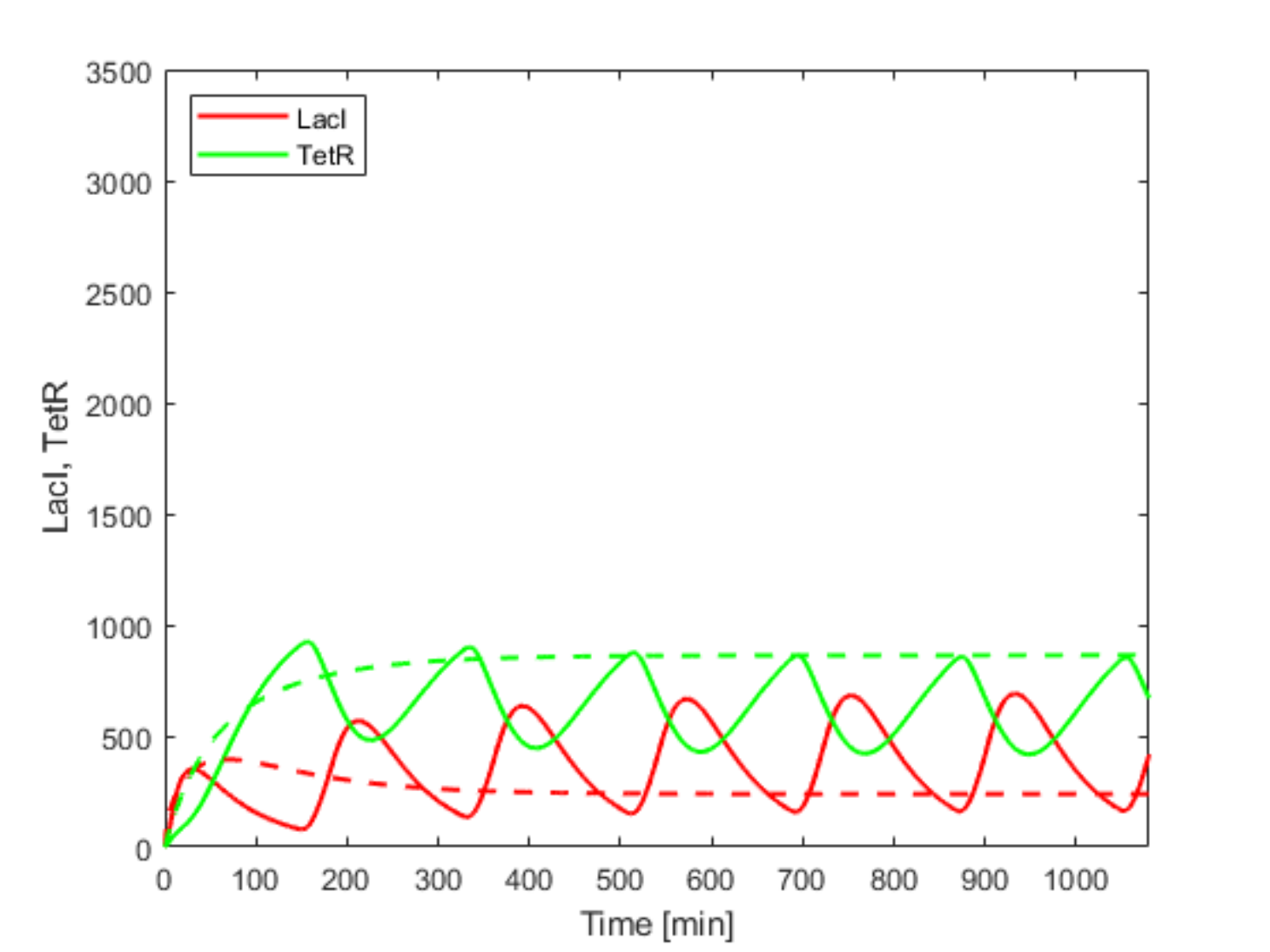}
\label{fig:tplot_atc050_iptg050_d080_diff}
}
\caption{Time evolution of the time-varying system (19) (in solid lines) and of the average system (20) (dashed lines) from initial conditions $[1,\, 1]^\mathsf{T}$ with $\bar{u}_\mathrm{aTc}=50\, \mathrm{ng/ml}$, $\bar{u}_{\mathrm{IPTG}}=0.5\, \mathrm{mM}$, $T=180\,\mathrm{min}$, $D=0.8$.}
\label{fig:tplot_d080}
\end{figure}

%%%%%%%%%%%%%%%%%%%%%%%%%%

\begin{figure}[!h]
\centering
\subfigure[With $T=180\,\mathrm{min}$, that is $\varepsilon \approx 3$.]
{
\includegraphics[width=0.4\linewidth]{tplot_atc050_iptg050_d050_no_diff}
\label{fig:tplot_atc050_iptg050_d050_no_diff_T_180}
}
\subfigure[With $T=45\,\mathrm{min}$, that is $\varepsilon\approx 0.75$]
{
\includegraphics[width=0.4\linewidth]{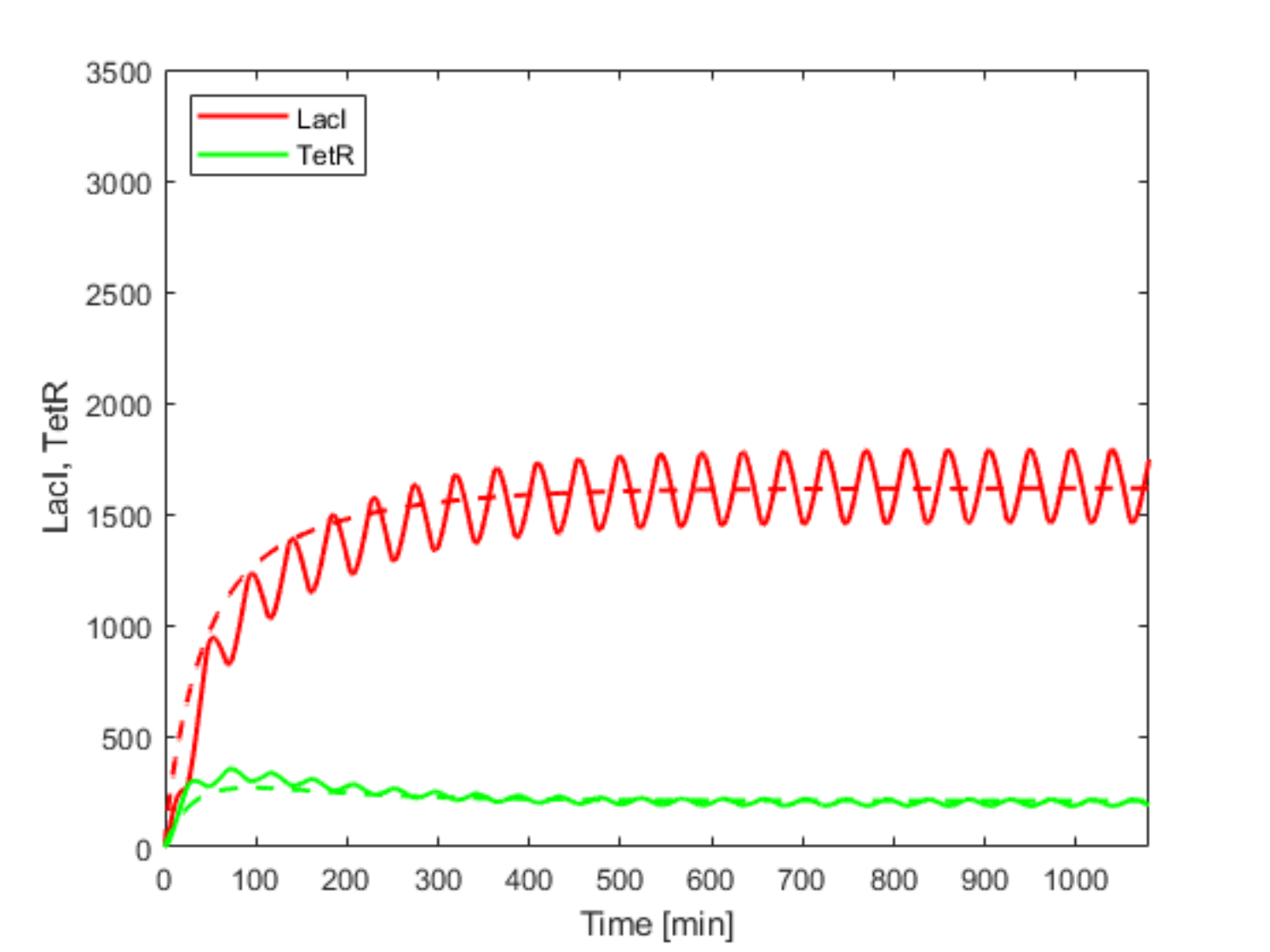}
\label{fig:tplot_atc050_iptg050_d050_no_diff_T_45}
}
\caption{Time evolution of the time-varying system (19) (in solid lines) and of the average system (20) (dashed lines) from initial conditions $[1,\, 1]^\mathsf{T}$ with $\bar{u}_\mathrm{aTc}=50\, \mathrm{ng/ml}$, $\bar{u}_{\mathrm{IPTG}}=0.5\, \mathrm{mM}$, $D=0.5$, with instantaneous diffusion (Assumption 1).}
\label{fig:tplot_d050_epsilon}
\end{figure}

%%%%%%%%%%%%%%%%%%%%%%%%%%%%%%%%%%%%%%%%%%%%%%%%%%%%%%%%%%%%%%%%%%%%%%%%%%%%

\begin{table}[!h]
\centering
%\footnotesize
\caption{Parameters of the toggle switch model (as reported in \cite{lugagne2017balancing})}
\bgroup
\def\arraystretch{1.5}
\begin{tabular}{|c|c|c|}
\hline
\multirow{4}{*}{\begin{tabular}[c]{@{}c@{}}Transcription rates\\ (mRNA min$^{-1}$)\end{tabular}} & $\kappa_\mathrm{L}^\mathrm{m0}$ & $3.20 \cdot 10^{-2}$ \\ \cline{2-3} 
 & $\kappa_\mathrm{T}^\mathrm{m0}$ & $1.19 \cdot 10^{-1}$ \\ \cline{2-3} 
 & $\kappa_\mathrm{L}^\mathrm{m}$ & $8.30$ \\ \cline{2-3} 
 & $\kappa_\mathrm{T}^\mathrm{m}$ & $2.06$ \\ \hline
\multirow{2}{*}{\begin{tabular}[c]{@{}c@{}}Translation rates\\ (a.u. mRNA$^{-1}$ min$^{-1}$)\end{tabular}} & $\kappa_\mathrm{L}^\mathrm{p}$ & $9.726 \cdot 10^{-1}$ \\ \cline{2-3} 
 & $\kappa_\mathrm{T}^\mathrm{p}$ & $1.170$ \\ \hline
\multirow{4}{*}{\begin{tabular}[c]{@{}c@{}}Degradation rates\\ (min$^{-1}$)\end{tabular}} & $g_\mathrm{L}^\mathrm{m}$ & $1.386 \cdot 10^{-1}$ \\ \cline{2-3} 
 & $g_\mathrm{T}^\mathrm{m}$ & $1.386 \cdot 10^{-1}$ \\ \cline{2-3} 
 & $g_\mathrm{L}^\mathrm{p}$ & $1.65 \cdot 10^{-2}$ \\ \cline{2-3} 
 & $g_\mathrm{T}^\mathrm{p}$ & $1.65 \cdot 10^{-2}$ \\ \hline
\multirow{4}{*}{\begin{tabular}[c]{@{}c@{}}\emph{plac} regulation\\ by LacI\end{tabular}} & $\theta_\mathrm{LacI}$ & $31.94$ a.u. \\ \cline{2-3} 
 & $\eta_\mathrm{LacI}$ & $2$ \\ \cline{2-3} 
 & $\theta_\mathrm{IPTG}$ & $9.06 \cdot 10^{-2}$ mM \\ \cline{2-3} 
 & $\eta_\mathrm{IPTG}$ & $2$ \\ \hline
\multirow{4}{*}{\begin{tabular}[c]{@{}c@{}}\emph{ptet} regulation\\ by TetR\end{tabular}} & $\theta_\mathrm{TetR}$ & $30.00$ a.u. \\ \cline{2-3} 
 & $\eta_\mathrm{TetR}$ & $2$ \\ \cline{2-3} 
 & $\theta_\mathrm{aTc}$ & $11.65$ ng/ml \\ \cline{2-3} 
 & $\eta_\mathrm{aTc}$ & $2$ \\ \hline
\multirow{2}{*}{\begin{tabular}[c]{@{}c@{}}IPTG exchange rate\\ (min$^{-1}$)\end{tabular}} & $k^{\mathrm{in}}_{\mathrm{IPTG}}$ & $2.75 \cdot 10^{-2}$ \\ \cline{2-3} 
 & $k^{\mathrm{out}}_{\mathrm{IPTG}}$ & $1.11 \cdot 10^{-1}$ \\ \hline
\multirow{2}{*}{\begin{tabular}[c]{@{}c@{}}aTc exchange rate\\ (min$^{-1}$)\end{tabular}} & $k^{\mathrm{in}}_{\mathrm{aTc}}$ & $1.62 \cdot 10^{-1}$ \\ \cline{2-3} 
 & $k^{\mathrm{out}}_{\mathrm{aTc}}$ & $2.00 \cdot 10^{-2}$ \\ \hline
\end{tabular}
\egroup
\end{table}


%%%%%%%%%%%%%%%%%%%%%%%%%%%%%%%%%%%%%%%%%%%%%%%%%%%%%%%%%%%%%%%%%%%%%%%%%%%%%%%%

%\bibliographystyle{IEEEtran} 
%\bibliography{refs} 

\end{document}